\documentclass[a4paper,11pt]{article}
\usepackage{jheppub}
\usepackage{lineno}

\usepackage{tikz}
\usetikzlibrary{positioning}
\usepackage{subcaption}
\usepackage{braket}

\arxivnumber{}

\makeatletter
\newcommand{\@fpheader}{}
\makeatother

\title{Meson thermalization with a hot medium in the open Schwinger model}

\author[1,2]{Takis Angelides \orcidlink{0000-0002-8639-8050},}
\author[2]{Yibin Guo \orcidlink{0000-0003-0435-1476},}
\author[3,2]{Karl Jansen \orcidlink{0000-0002-1574-7591},}
\author[2]{Stefan Kühn \orcidlink{0000-0001-7693-350X},}
\author[4,5]{Giuseppe Magnifico \orcidlink{0000-0003-2376-5682}}

\affiliation[1]{Institut für Physik, Humboldt-Universität zu Berlin,\\Newtonstr. 15, 12489 Berlin, Germany}
\affiliation[2]{Deutsches Elektronen-Synchrotron DESY,\\Platanenallee 6, 15738 Zeuthen, Germany}
\affiliation[3]{Computation-Based Science and Technology Research Center, The Cyprus Institute,\\20 Kavafi Street, 2121 Nicosia, Cyprus}
\affiliation[5]{Dipartimento di Fisica, Università di Bari,\\I-70126 Bari, Italy}
\affiliation[6]{Istituto Nazionale di Fisica Nucleare (INFN), Sezione di Bari,\\I-70125 Bari, Italy}

\emailAdd{takis.angelides@desy.de}

\date{\today}

\abstract{Quantum field theories treated as open quantum systems provide a crucial framework for studying realistic experimental scenarios, such as quarkonia traversing the quark-gluon plasma produced at the Large Hadron Collider. In such cases, capturing the complex thermalization process requires a detailed understanding of how particles evolve and interact with a hot medium. Considering the open lattice Schwinger model and using tensor network algorithms, we investigate the thermalization dynamics of mesonic particles in a hot medium, such as the Schwinger boson or the electric flux string. We simulate systems with up to 100 lattice sites, achieving accurate preservation of the electric field parity symmetry, demonstrating the algorithm's robustness and scalability. Our results reveal that the thermalization time increases with stronger dissipation from the environment, increasing environment temperature, higher background electric field and heavier fermion masses. Further, we study the quantum mutual information between the two halves of the flux string connecting a meson's constituent particles and analyze its relation to relevant dynamical observables.}

\begin{document}
\maketitle
\flushbottom

\section{Introduction}

Open quantum systems (OQS) describe scenarios where a quantum system of interest interacts with an environment, influencing system dynamics through exchanges of energy, particles, and information~\cite{Breuer:2002pc, Minganti:2024uic, Breuer2003, takahashi, bharath}. Since all experiments investigating quantum phenomena inherently involve open systems, and nature is effectively described by quantum field theories (QFT)~\cite{Schwartz:2014sze}, the development and analysis of open quantum field theories are of paramount importance. For example, heavy-ion collisions at the Large Hadron Collider or the Relativistic Heavy Ion Collider create a hot, dense medium known as the quark-gluon plasma (QGP), within which bound heavy quark states called quarkonia (e.g. the bottomonium or charmonium), can propagate~\cite{qgp_review, RAPP2010209, oqs_for_quarkonia, refId0}. In this case, the environment is the QGP, and the system is the quarkonium~\cite{He:2022ywp}. Quarks within the QGP are believed to experience color-charge screening~\cite{Mocsy_Petreczky_Strickland_2013}, similar to electric charge screening in $1+1$ dimensional quantum electrodynamics~\cite{COLEMAN1975267}. This screening may reduce the yield of bound quarkonia~\cite{MATSUI1986416}, making quarkonia an effective probe for studying QGP properties~\cite{Andronic, Mocsy_Petreczky_Strickland_2013, Blaizot, PhysRevD.108.054024}. The variation in mass across these bound states affects their sizes, influencing their screening, dissociation, and thermalization. This has been studied in~\cite{PhysRevD.101.034011, PhysRevD.101.074004} through master equation techniques, which enable an exact quantum state representation by restricting the Hilbert space dimension.

To explore meson thermalization within a hot medium, we employ the Schwinger model (SM) as a toy model for quantum chromodynamics (QCD), because of its shared QCD characteristics, such as confinement and charge screening~\cite{COLEMAN1975267, manton1985schwinger}. Embedded within a tensor network (TN) framework~\cite{tn_review, annurev:/content/journals/10.1146/annurev-conmatphys-040721-022705, Silvi_2014, Banuls2019TensorNetworksTheir, Review_Banuls_Cichy}, which has been successfully applied to the study of both equilibrium and dynamical properties of the Schwinger model~\cite{Buyens2014MatrixProductStates, Rico2014TensorNetworksLattice, PhysRevX.6.011023, PhysRevLett.118.071601, Ercolessi2018PhaseTransitionsZn, Magnifico2020RealTimeDynamics, PhysRevD.101.054507, Rigobello2021EntanglementGeneration11d, Angelides:2023bme, Funcke2023ExploringCpViolating, osborne2023probing, PhysRevLett.132.091903, Papaefstathiou2024realtime, jeyaretnam2024hilbertspacefragmentationorigin, guo2024concurrentvqesimulatingexcited}, we overcome traditional Monte Carlo sign problems~\cite{sign_problem} and the limitations of perturbative methods at strong couplings.

For our numerical studies, the space is discretized into a finite lattice of $N$ sites, yielding the lattice SM, which consists of fermionic degrees of freedom (DOF) on sites and gauge DOF on links between the sites. We couple this system to an environment represented by a $\phi^4$-theory at high temperature in thermal equilibrium, which is suitable for modeling the QGP~\cite{oqs_for_quarkonia}. In the Markovian quantum Brownian motion limit, we make use of the Lindblad master equation~\cite{Breuer:2002pc, oqs_for_quarkonia}, enabling us to simulate the time evolution of the system's density matrix and analyze its dynamics. To this end, we have developed TN code that leverages the adaptive time-dependent density matrix renormalization group (DMRG) algorithm~\cite{Xiang_2023} to simulate open quantum field theory dynamics. To the best of our knowledge, this work represents the first application of tensor network methods to simulate a lattice gauge theory within the framework of open quantum systems.

Simulating dynamics analogous to quarkonia in QGP, we examine string dynamics and the dynamics of the Schwinger boson in a hot environment, using the electric field as an observable to track thermalization influenced by temperature $T$, mass $m$, background electric field $l_0$, and dissipation strength $D$ of the environment on the system. We further analyze thermalization via mutual information from quantum information theory~\cite{nielsen2000}, offering insights consistent with experimental and numerical studies on quarkonia in QGP.

To date, there have been very few studies investigating the Schwinger model as an open quantum system. In~\cite{open_schwinger_original}, the authors perform a time evolution of the Dirac vacuum for $N = 2$ on an IBM quantum device with fixed parameters and use a simulator to increase $N$ up to 8 lattice sites, analyzing finite-size effects. String breaking in open vs. closed systems is explored in~\cite{open_schwinger}, highlighting the differences in dynamics between isolated and dissipative environments. Lastly,~\cite{neural_nets_oqs_schwinger} presents a method using neural density operators to represent the density matrix with neural networks, simulating string dynamics for up to $N = 32$ and three interacting strings.

Despite these advances, several open questions remain regarding the impact of gauge DOF truncation, the role of external fields in string dynamics, and the dynamics of the theory's stable particle, called the Schwinger boson. Prior studies often truncate the gauge DOF Hilbert space to allow only a small number of states per link, potentially limiting the proximity to the continuum limit. Additionally, the influence of an applied background electric field on string behavior in open quantum systems (OQS) has not been explored.

Our work provides a step towards addressing these open questions by studying an open lattice Schwinger model without truncating the gauge DOF. Instead, we replace the gauge DOF with fermionic DOF by using open boundary conditions and Gauss's law, thereby avoiding truncation entirely. Further, we investigate the Schwinger boson, a stable meson particle of the theory~\cite{Coleman:1976uz, ADAM1996111}, and introduce an applied background electric field, shown to impact string dynamics in the OQS. This feature could have practical implications for future experimental studies of quarkonia in the quark-gluon plasma (QGP). Finally, by simulating system sizes up to $N = 100$, we demonstrate that our method can scale to larger lattice sizes, an essential step towards bridging numerical and experimental results in the continuum limit of quantum field theory.

The paper is organized as follows. The theoretical framework for the open lattice Schwinger model is described in section~\ref{sec:theory}. We proceed with the explanation of our time evolution scheme and the tensor network ansatz in section~\ref{sec:methods}. Results are included in section~\ref{sec:results_discussion} which breaks down to the following subsections. In subsections~\ref{sec:thermalization_times_vs_various_parameters},~\ref{sec:mutual_info} and~\ref{sec:temperature} we work with the electric flux string meson and respectively present results on how $D, l_0$, $m$ affect the thermalization time $\mathcal{T}$, the correlations between mutual information and $\mathcal{T}$, and how temperature influences the mutual information and thermalization time. Moving to subsection~\ref{sec:larger_systems}, results of larger system sizes up to 100 lattice sites are presented and results for the Schwinger boson analysis are shown in subsection~\ref{sec:schwinger_boson_results}. A conclusion and outlook are given in section~\ref{sec:conclusion}, while the two appendices~\ref{app:density_matrix_as_mps} and~\ref{app:atddmrg} provide respectively information on the density matrix representation with MPS and the time evolution algorithm we use.

\section{Theoretical framework of the open lattice Schwinger model}
\label{sec:theory}

An open quantum system is described by the system of interest $S$ and its environment $E$ through the Hamiltonians $H_S$, $H_E$ and $H_I$, where the latter is the interaction Hamiltonian between $S$ and $E$~\cite{Breuer:2002pc, Rivas:2012ugu}. It is usually the case that we are interested in measuring observables only on $S$ but nevertheless want to study the effect of $E$ on $S$. Hence, we only aim to follow the dynamics of the reduced density matrix for the system $\rho_S$ defined by tracing out the degrees of freedom of $E$. 

For $H_S$ we take the Schwinger model with a background electric field $E_0 = g\theta/2\pi \equiv gl_0$ and one fermion flavour represented by the two-component spinor $\psi$. In the continuum limit, using the temporal gauge $A_0 = 0$, $H_S$ has the form~\cite{Angelides:2023noe}
\begin{align}
\begin{split}
\label{eq:H_S_continuum}
    H_S &= \int dx \left[-i\overline{\psi}\gamma^1\left(\partial_1 - igA_1\right)\psi + m\overline{\psi}\psi + \frac{1}{2}\left(\dot{A}_1 + \frac{g\theta}{2\pi}\right)^2\right].
\end{split}
\end{align}
where $A_1$ is the spatial component of the gauge field $A_\mu, \mu \in [0, 1]$. Here, $m$ is the bare fermion mass and $g$ is the bare coupling between the fermions and the gauge field. Physical eigenstates of $H_S$ also need to obey the constraint of Gauss's law $\partial_1 \dot{A}_1 = g\psi^\dagger \psi$.

For numerical calculations, it is necessary to discretize Eq.~\eqref{eq:H_S_continuum} on a lattice of discrete spatial sites. To avoid the doubling problem~\cite{NIELSEN1981219, Gattringer:2010zz}, we use the staggered formulation~\cite{KS_hamiltonian_formulation}
\begin{align}
\label{eq:staggered_hamiltonian_initial}
    \begin{aligned}
        H_S &= -\frac{i}{2a}\sum_{n=0}^{N-2} \left(\chi^\dagger_n U_n\chi_{n+1}-\text{h.c}.\right) + m_\text{lat}\sum_{n=0}^{N-1} (-1)^n\chi^\dagger_n\chi_n + \frac{ag^2}{2}\sum_{n=0}^{N-2} \left(l_0 + L_n\right)^2.
    \end{aligned}
\end{align}
In the above, $n$ is the index enumerating the spatial lattice sites, $N$ is the total number of sites, and $a$ is the lattice spacing between them. We now have a single component fermionic field $\chi$ and the link operators $U_n$ which are the ladder operator of the electric field. The notation $U_n$ then implies that this operator is placed on the link to the right of site $n$ and ensures local gauge invariance. The mass term explicitly distinguishes the bare continuum mass $m$ from the bare lattice mass $m_{\text{lat}}$ which are related by a constant shift in a closed system setting ~\cite{Angelides:2023bme}. For simplicity, for the rest of this paper we will refer to $m_{\text{lat}}$ as $m$. The last term involves the dimensionless electric field operator $L_n = E_n/g$ that has an infinite-dimensional Hilbert space, and the background electric field in units of the coupling $l_0 = \theta/2\pi$.

Physical states need to obey the discrete version of Gauss's law given by
\begin{equation}
    L_n - L_{n-1} = Q_{n} + Q_{n}^{\text{ext}},
\end{equation}
where $Q_n = \chi_n^\dagger \chi_n - (1-(-1)^n)/2$ is the charge operator and $Q_n^{\text{ext}}$ the external charge operator on the site $n$. Since we do not use external charges the latter can be omitted. 

For open boundary conditions (OBC), which we use in our simulations, the above can be solved iteratively after fixing the left most electric field value to zero. The solution to Gauss's law is then
\begin{equation}
\label{eq:electric_field_operator}
    L_n = \sum_{k = 0}^{n-1}Q_k,
\end{equation}
which combined with a unitary transformation on the fermion fields~\cite{hamer97} allows for eliminating the gauge fields $U_n, L_n$ completely up to $l_0$, substituting them with the fermionic fields $\chi$.

It is then convenient to work with a dimensionless Hamiltonian and operators expressed in terms of Pauli spin operators achieved by rescaling and using the Jordan-Wigner transformation $\chi_n = \prod_{k<n} (iZ_k)\sigma^-_n$~\cite{jordan_wigner}  respectively. The dimensionless Hamiltonian $aH_S$ on the lattice is then given by~\cite{Angelides:2023noe}
\begin{align}
\begin{split}
\label{eq:H_S}
    aH_S &= x\sum_{n=0}^{N-2} \left(S^+_nS^-_{n+1} + S^-_nS_{n+1}^+\right) \\
    &\quad +\frac{1}{2}\sum_{n=0}^{N-2}\sum_{k=n+1}^{N-1}(N-k-1)Z_nZ_k\\
    &\quad +\sum_{n=0}^{N-2}\left(\frac{N}{4}-\frac{1}{2}\left\lceil\frac{n}{2}\right\rceil+l_0(N-n-1)\right)Z_n \\
    &\quad +\frac{m_{\text{lat}}}{g}\sqrt{x}\sum_{n=0}^{N-1} (-1)^n Z_n\\
    &\quad +l_0^2(N-1) + \frac{1}{2}l_0N + \frac{1}{8}N^2.
\end{split}
\end{align}
The first term is the kinetic hopping term which defines the inverse of the lattice spacing $a$ in units of the coupling $g$ between the fermions and gauge fields as $x \equiv 1/(ag)^2$. In this term $S_n^{+,-}$ are the spin-1/2 ladder operators on site $n$. As aforementioned, the physical states need to obey Gauss's law and eigenstates of the above Hamiltonian automatically obey this constraint, as we have solved explicitly Gauss's constraint in Eq.~\eqref{eq:electric_field_operator} and substituted it in to eliminate the gauge fields. This is the origin of the second term, which is a long-range interaction stemming from the Coulomb force. The $Z_n$ is the Pauli $Z$ operator on site $n$. The third term of Eq.~\eqref{eq:H_S} is part of the electric field energy, where $l_0$ is defined above as the dimensionless background electric field. Next we have the mass term where $m/g$ is the lattice mass in units of the coupling. The last term involves constants coming from the previous terms. For further details of this derivation see~\cite{Angelides:2023noe}. 

We note here that the charge operator after the Jordan-Wigner transformation becomes
\begin{equation}
\label{eq:charge_operator}
    Q_n = \frac{Z_n+(-1)^n}{2}.
\end{equation}
Another important note, which we will make use of in Sec.~\ref{sec:results_discussion}, is the description of the ground state of Eq.\eqref{eq:H_S} in the heavy mass limit. In this limit, and assuming the background electric field is set to zero, the ground state has no particle excitations and no electric field. Given Eq.~\eqref{eq:charge_operator} and the fourth line of Eq.~\eqref{eq:H_S}, we can see that this ground state, also called the Dirac vacuum, is given by the product state $\ket{01..01}$ and is absent of any charges.

Regarding $H_E$ and $H_I$, we follow~\cite{open_schwinger} in which the environment is assumed to be in thermal equilibrium at temperature $T = 1/\beta$ for all times described by a scalar $\phi^4$-theory, and $H_I$ is a Yukawa interaction
\begin{align}
    &H_E = \int dx \left[\frac{1}{2}\Pi^2 + \frac{1}{2}\left(\nabla \phi\right)^2 + \frac{1}{2} m^2_{\phi}\phi^2 + \frac{1}{4!}g\phi^4\right] \\
    &H_I = \int dx \lambda \phi(x)\bar\psi(x) \psi(x),
\end{align}
with $\lambda$ determining the strength of the interaction between the system and the environment. 

Working in the Markovian limit, we assume that the interaction between the system and the environment is weak such that the density matrix can be approximated to be in the product state
\begin{equation}
    \rho(t) = \rho_S(t) \otimes \rho_E,
\end{equation}
where $\rho_E = e^{-\beta H_E}/ \text{Tr}(e^{-\beta H_E})$ is the environment's density matrix in the Gibbs state. Further, we work in the quantum Brownian motion (QBM) limit given by the separation of time scales $\tau_R \gg \tau_E$, $\tau_S \gg \tau_E$. The first inequality implies that the system with relaxation time $\tau_R \sim T/H_I^{(\text{int})^2}$ relaxes much more slowly than the environment's correlation time $\tau_E \sim 1/T$, where $H_I^{(\text{int})}$ is the interaction Hamiltonian in the interaction picture. We can interpret $H_I^{(\text{int})}$ in the definition of $\tau_R$ as the gap between the ground and first excited state of $H_I^{(\text{int})}$. The second inequality suggests that the intrinsic time scale of the system $\tau_S \sim 1/H_S$ is much greater than $\tau_E$, which is valid in the limit $T \gg H_S$. Here we can interpret $H_S$ to be the spectral gap of $H_S$ which is $1/\sqrt{\pi}$ at $m = 0$ in the continuum limit~\cite{adam}. These limits further justify the assumption that $\rho_E$ remains for all times in the thermal Gibbs state defined by $H_E$ as any perturbation on the environment's Gibbs state will decay too fast to affect the system which we are interested in. The high temperature limit we are using, where the environment is always in thermal equilibrium, is relevant for the case of a hot nuclear QGP medium~\cite{oqs_for_quarkonia}. Under these assumptions, the evolution of the density matrix of $S$ is given by the Lindblad master equation
\begin{equation}
\label{eq:lme}
    \frac{d\rho_S(t)}{dt} = -i\left[H_S, \rho_S(t)\right] + a^2 \sum_{n, k = 0}^{N-1} D(n - k) \left(J(k)\rho_S(t) J^\dagger(n) - \frac{1}{2}\left\{J^\dagger(n)J(k),\rho_S(t)\right\}\right),
\end{equation}
in which $D(n-k)$ is the environment correlator
\begin{equation}
    D(n-k) = \lambda^2 \int_{-\infty}^{\infty} dt_1 \int_{-\infty}^{\infty}dt_2\text{Tr}_E\left[\phi^{(\text{int})}(t_1,n)\phi^{(\text{int})}(t_2,k)\rho_E\right].
\end{equation}

A detailed derivation of the above can be found in~\cite{oqs_for_quarkonia}. The fields $\phi^{(\text{int})}$ are in the interaction picture and $D(n-k)$, which is also called the dissipator, only depends on the distance between the two spatial points on the lattice. Finally, the Lindblad jump operators are defined as
\begin{align}
\label{eq:lindblad_jump_operator_definition}
    &J(n) = O(n) - \frac{1}{4T}\left[H_S, O(n)\right] \\
    &O(n) = (-1)^n\frac{Z_n+1}{2a}.
\end{align}
The next section will explain the methods used to simulate the time evolution of Eq.~\eqref{eq:lme} using Trotterization and a tensor network ansatz.

\section{Lindbladian evolution scheme with tensor networks}
\label{sec:methods}

In this section, we discuss the methods used to evolve the density matrix $\rho_S(t)$ under the Lindblad master equation. We describe how $\rho_S(t)$ is expressed as a matrix product state and the integration scheme we use in Trotterization for the time evolution.  

We can multiply Eq.~\eqref{eq:lme} by $a$ to express all terms in dimensionless form
\begin{align}
\begin{split}
\label{eq:lme_dimless}
    \frac{d\rho_S(t/a)}{d(t/a)} = &-i\left[aH_S, \rho_S(t/a)\right] \\
    &+ \sum_{n, k = 0}^{N-1} aD(n - k) \left(aJ(k)\rho_S(t/a) aJ^\dagger(n) - \frac{1}{2}\left\{aJ^\dagger(n)aJ(k),\rho_S(t/a)\right\}\right).
\end{split}
\end{align}
For the rest of this paper, all variables and operators are expressed in their dimensionless form in units of the lattice spacing $a$, and we suppress the lattice spacing.

The evolution of the density matrix $\rho_S(t)$ under the Lindblad master equation Eq.~\eqref{eq:lme_dimless}, which can be written as $\dot \rho_S(t) = \mathcal{L}\rho_S(t)$, has the formal solution
\begin{equation}
    \rho_S(t) = e^{t\mathcal{L}}\rho_S(t=0),
\end{equation}
where we can define the Liouvillian superoperator $\mathcal{L}$ that generates the dynamics as
\begin{align}
\begin{split}
\label{eq:lindblad_operator_double_space_form}
    \mathcal{L} = &-i H_S \otimes I + i I \otimes H_S \\
    &+ \sum_{n, k = 0}^{N-1} D(n-k)\left(J(k)\otimes J^\dagger(n) - \frac{1}{2}J^\dagger(n)J(k)\otimes I - \frac{1}{2}I \otimes J^\dagger(n)J(k)\right).
\end{split}
\end{align}
Formally, $\rho_S(t)$ can be represented as a matrix product operator (MPO), a type of tensor network that describes operators~\cite{vidal2004, orus2014}. The mathematical expression for an MPO of $N$ spatial sites with OBC is a product of rank-3 and rank-4 tensors $W$ given by
\begin{align}
\begin{split}
\label{eq:mpo_main_text}
    &\rho_S(t)_{\sigma_0...\sigma_{N-1}}^{\sigma'_0...\sigma'_{N-1}}\ket{\sigma'_0...\sigma'_{N-1}}\bra{\sigma_0...\sigma_{N-1}} \\
    &=W^{\sigma'_0}_{\sigma_0\alpha_0}W^{\sigma'_1}_{\sigma_1\alpha_0 \alpha_1}...W^{\sigma'_{N-1}}_{\sigma_{N-1}\alpha_{N-2}}\ket{\sigma'_0...\sigma'_{N-1}}\bra{\sigma_0...\sigma_{N-1}},
\end{split}
\end{align}
where $\sigma \in [0, 1]$ are the physical indices, which are implicitly summed, and represent the spin up or down degrees of freedom on the spatial sites. The $\alpha$ are implicitly summed bond indices, which have a dimension $D$ called the bond dimension. This controls the size of the tensors $W$ and further controls how much entanglement exists between any two sites~\cite{vidal2007}. The meaning of the tensor product in Eq.~\eqref{eq:lindblad_operator_double_space_form} can now be explained by Eq.~\eqref{eq:mpo_main_text}, where operators on the left side of the product act on the $\sigma'$ and on the right side act on the $\sigma$ indices. We reshape our density matrix into an MPS via singular value decomposition (SVD), in a similar manner to how a matrix can be reshaped into a vector. This procedure is explained in detail in appendix~\ref{app:density_matrix_as_mps}. Using again the implicit Einstein summation over the physical indices $\sigma$ and bond indices $\alpha$, we can express the MPS corresponding to Eq.~\eqref{eq:mpo_main_text} as a product of rank-2 and rank-3 tensors $A$, given by
\begin{align}
\begin{split}
    \label{eq:density_matrix_from_mpo_to_mps_methods}
    &\rho_S(t)_{\sigma'_0\sigma_0...\sigma'_{N-1}\sigma_{N-1}}\ket{\sigma'_0\sigma_0...\sigma'_{N-1}\sigma_{N-1}} \\ &=A_{\sigma'_0 \alpha_0} A_{\sigma_0\alpha_0\alpha_1}...A_{\sigma'_{N-1} \alpha_{2N-2}\alpha_{2N-1}} A_{\sigma_{N-1}\alpha_{2N-1}}\ket{\sigma'_0\sigma_0...\sigma'_{N-1}\sigma_{N-1}}.
\end{split}
\end{align}
From Eq.~\eqref{eq:density_matrix_from_mpo_to_mps_methods} we can see that the $\sigma'$ indices are on the even sites and the $\sigma$ indices on the odd sites. Given the above, this means the operators on the left side of the tensor products in Eq.~\eqref{eq:lindblad_operator_double_space_form} act on the even sites and the operators on the right side of the tensor products act on the odd sites. More details on this, for example on how to measure relevant observables, are given in appendix~\ref{app:density_matrix_as_mps}. 

The terms in the operator $\mathcal{L}$ of Eq.~\eqref{eq:lindblad_operator_double_space_form} are then partitioned in three groups of what we call even, odd, Taylor. The even and odd are groups of operators spanning four sites. The Taylor group encompasses all other operators that cannot be part of the even or odd groups. For example, the Taylor group includes interactions that can span the whole lattice coming from terms such as the second line of Eq.~\eqref{eq:H_S}. For a more explicit explanation on this partition, we refer the reader to appendix~\ref{app:atddmrg}. The even and odd groups are represented exactly as rank 8 tensors, whereas the Taylor group is Taylor expanded to order $\kappa$ as
\begin{equation}
\label{eq:taylor_group_taylor_expansion}
    e^{\tau \mathcal{L}_{T}} \approx 1 + \sum_{j=1}^{\kappa}\frac{\left(\tau \mathcal{L}_T\right)^j}{j!},
\end{equation}
where $\mathcal{L}_T$ is the Taylor group of terms in $\mathcal{L}$. If we let $\mathcal{L}_E$ and $\mathcal{L}_O$ be the corresponding even and odd groups, then our second order Trotterization scheme approximates $e^{\tau\mathcal{L}}$ as
\begin{equation}
\label{eq:Trotterization}
    e^{\tau\mathcal{L}} \approx e^{\frac{\tau}{2}\mathcal{L}_E}e^{\frac{\tau}{2}\mathcal{L}_T}e^{\tau\mathcal{L}_O}e^{\frac{\tau}{2}\mathcal{L}_T}e^{\frac{\tau}{2}\mathcal{L}_E} + \mathcal{O}\left(\tau^2\right).
\end{equation}

We represent the right-hand side of Eq.~\eqref{eq:taylor_group_taylor_expansion} as a global MPO with two singular value cutoffs $\epsilon_1$, $\epsilon_2$. Using the ITensors Julia library~\cite{ITensors} allows to convert $\mathcal{L}_T$ automatically to an MPO and can apply a truncation on the singular values in the process, which we call $\epsilon_1$. The ratio of the sum of squares of the neglected singular values to the sum of squares of all singular values will not exceed this cutoff at any given singular value decomposition truncation. Then $\epsilon_2$ is used when we are multiplying $\mathcal{L}_T$ onto itself to form powers of this operator as needed for the right hand side of Eq.~\eqref{eq:taylor_group_taylor_expansion}. Finally, we use the adaptive time-dependent DMRG algorithm~\cite{Xiang_2023} to apply the above operators with a singular value cutoff $\epsilon$, and evolve our state $\rho_S(t)$. For more details on this algorithm and the partition of $\mathcal{L}$ into groups see appendix~\ref{app:atddmrg}. Throughout this paper we fix $D(n-k) = D\delta_{n,k}$~\cite{open_schwinger_original}, $x = 1$, $\epsilon = 10^{-11}$, $\epsilon_1 = \epsilon_2 = 10^{-9}$, $\kappa = 2$ and $\tau = 0.01$, except from section~\ref{sec:larger_systems} where we set $x = 4$, $\tau = 0.001$ due to the larger $N$ used. For the high-temperature limit in which we are working, the environment's correlation length is well approximated by $1/T$, implying that it is small. Hence, the delta function dissipator provides a good description for the environment in this regime.

The above methods allow us to simulate the time evolution of meson type states according to Eq.~\eqref{eq:lme_dimless} and gather results for their dynamics in a hot medium which we turn to in the next section.

\section{Meson thermalization dynamics}
\label{sec:results_discussion}

In this section we will present and discuss the results from the time evolution of two types of meson initial states. One of these states is created from the Dirac vacuum which is the state in the absence of charge. This is the eigenstate of the Hamiltonian in Eq.~\eqref{eq:H_S} when the infinite mass limit is taken and can be easily prepared with the product state $\ket{01..01}$, as briefly discussed in Sec.~\ref{sec:theory}. We then introduce a string of electric flux onto this state by flipping the spins at sites $N/2 - 2$ and $N/2 + 1$, where counting begins from 0. This creates a positive unit charge on site $N/2 - 2$ and a negative one on site $N/2 + 1$ which is easily seen from Eq.~\eqref{eq:charge_operator}. We can then evolve this string under the influence of the hot environment and track relevant observables. To isolate the dynamics of the string, we additionally perform the time evolution of the Dirac vacuum state and subtract its corresponding observable values from the observable values of the string state. We refer to these observables as subtracted observables. For example, an important observable is the electric field as a function of the link number which is given by the expectation value of the operator in Eq.~\eqref{eq:electric_field_operator} $F(n) \equiv \langle L_n \rangle$, and we thus symbolize the subtracted electric field (SEF) as $\Delta F(n)$. 

\begin{figure}
    \centering
    \includegraphics[width=\linewidth]{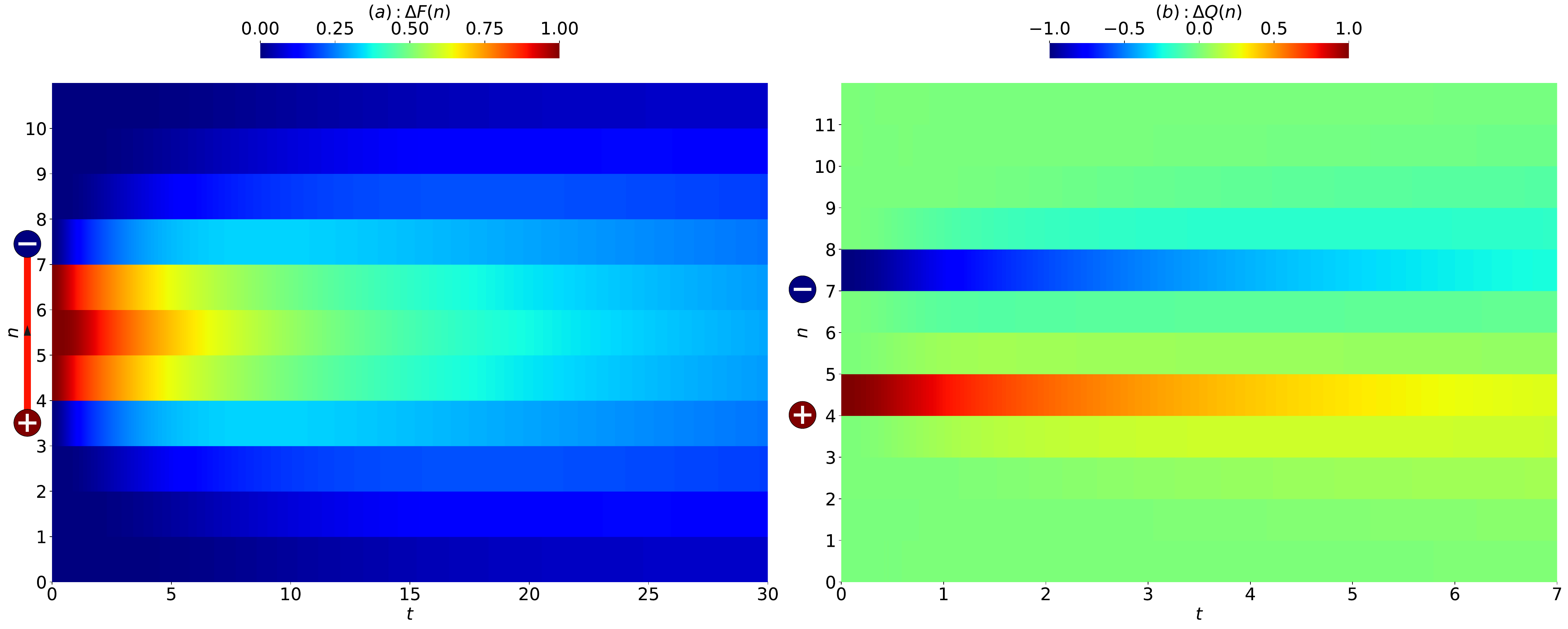}
    \caption{$(a)$: Subtracted electric field $\Delta F(n)$ per link $n$ as a function of time $t$. The red line on the y-axis represents the initial electric field flux generated by the pair of positive/negative charges shown as plus/minus on the string's endpoints. $(b)$: Subtracted charge $\Delta Q(n)$ per site $n$ as a function of time $t$. In $(b)$ we focus on the early time dynamics and show how the charges forming the initial string spread out. The parameters used are $N = 12$, $x = 1$, $m = 0.5$, $l_0 = 0$, $D = 2$, $T = 10$.}
    \label{fig:inkscape_graphic}
\end{figure}

In figure~\ref{fig:inkscape_graphic}$(a)$ we show an example of the SEF of the string introduced in the previous paragraph as a function of time using $N = 12$. We can observe the initial electric flux of the string and how it spreads with time and eventually thermalizes. Further, the middle link at $n = 5$ shows the largest thermalization time, a feature we will make use of to define a thermalization time. The corresponding subtracted charge $\Delta Q(n)$ is shown in figure~\ref{fig:inkscape_graphic}$(b)$ in which the initial unit charges are visible on the sites where we have flipped the spins of the Dirac vacuum. These charges then spread out with time as thermalization takes place and become less localized.

The second meson-type state we use as an initial state is the first excited state of the Hamiltonian $H_S$. In this case we subtract from observables the corresponding values of the ground state of $H_S$ which isolates the dynamics of the Schwinger boson. This is a stable meson particle of the theory with weak self-interactions~\cite{Coleman:1976uz}. To distinguish subtracted observables for this initial state, as opposed to the string case, we use the subscript $B$ and refer for example to the subtracted electric field as $\Delta F_B(n)$.

In section~\ref{sec:thermalization_times_vs_various_parameters} we look at the electric field of the string case as a guiding observable in defining the thermalization time and how the latter is affected by the mass $m$, the applied background electric field $l_0$ and the dissipation strength $D$. To further investigate the thermalization process, we present in section~\ref{sec:mutual_info} results of mutual information across the two halves of the string and in section~\ref{sec:temperature} how the string thermalization is affected by the temperature. Section~\ref{sec:larger_systems} demonstrates simulations with larger system sizes with the purpose of testing the thermalization time results of section~\ref{sec:thermalization_times_vs_various_parameters} against finite-size effects, exemplifying that the algorithm we use can perform simulations with sizes of $\mathcal{O}(100)$ sites and testing the precision at which the parity symmetry of the SEF about the middle link is preserved. This parity symmetry is a consequence of choosing $D(n-k) = D\delta_{n,k}$ and implies $L_n = L_{N-2-n}$~\cite{neural_nets_oqs_schwinger} as discussed in more detail in section~\ref{sec:larger_systems}. Finally, in section~\ref{sec:schwinger_boson_results} the results of the Schwinger boson are presented, specifically focusing on the effect of the dissipator strength of the environment on the Schwinger boson's thermalization time.

\subsection{\texorpdfstring{Effects of $D$, $l_0$ and $m$ on string thermalization}{Effects of D, l0 and m on string thermalization}}
\label{sec:thermalization_times_vs_various_parameters}

\begin{figure}
    \centering
    \includegraphics[width=\linewidth]{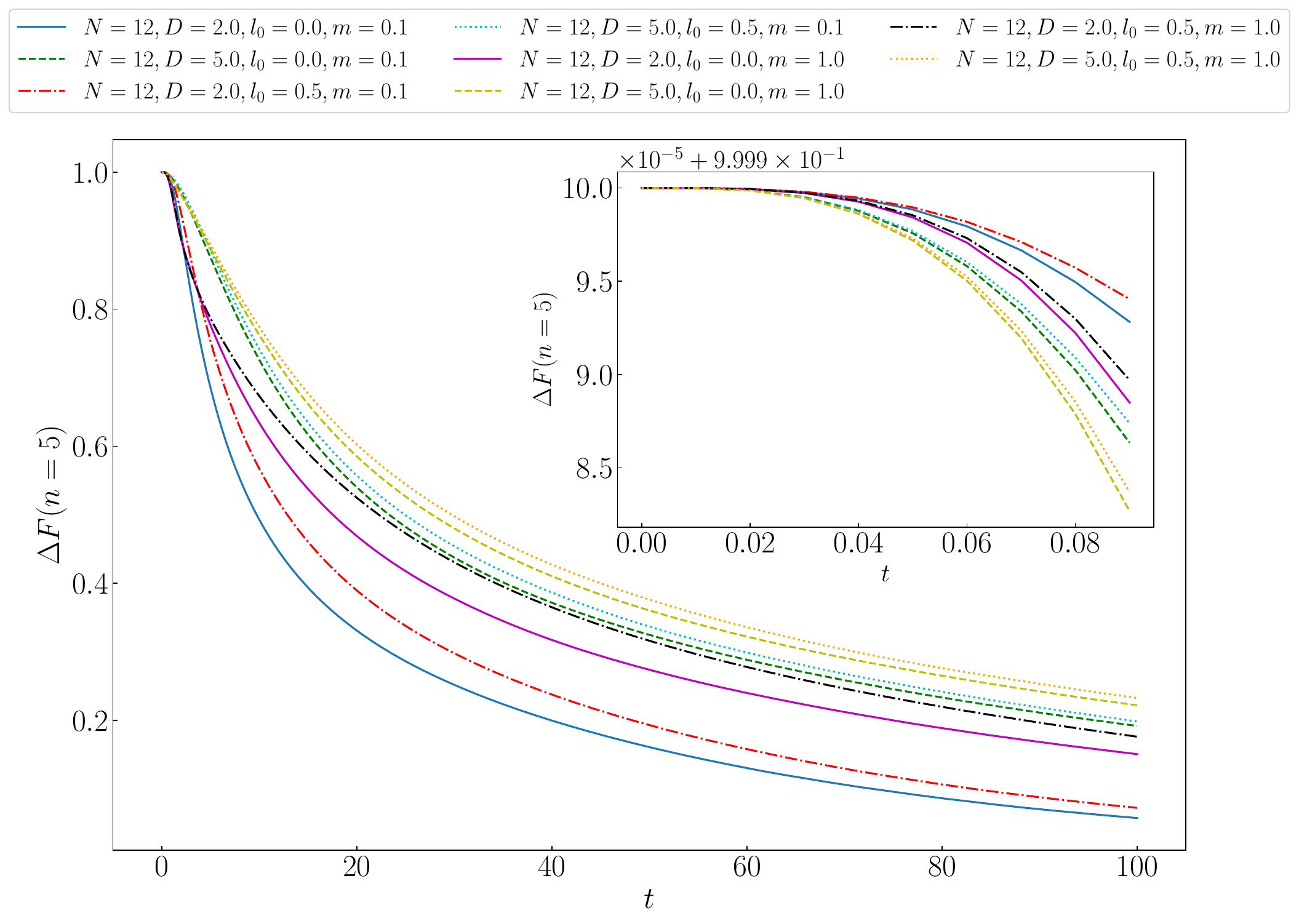}
    \caption{Subtracted electric field (SEF) $\Delta F(n)$ of the middle link $n = 5$ as a function of time $t$. The inset shows the early time dynamics and emphasizes the fact that SEF monotonically decreases from its initial value to zero towards the steady state. The resolution in $t$ is fine enough that the data is quasi-continuous and connected with lines. This monotonic behaviour of the SEF facilitates the definition of the thermalization time $\mathcal{T}$.}
    \label{fig:monotonicity_of_sef}
\end{figure}

Our goal for this subsection is to measure the thermalization time $\mathcal{T}$ of the string as the parameters $m$, $l_0$, and $D$ are varied. Both initial states—the Dirac vacuum and the one including the string—eventually reach the same steady state, as it is determined by the Lindblad operator, which is independent of the initial condition~\cite{open_schwinger_original, open_schwinger}. Consequently, when the SEF is zero it implies thermalization and we have observed empirically that the SEF goes to zero monotonically for all links at late times. It has further been observed that the middle link consistently exhibits the largest thermalization time. Thus, we define $\mathcal{T}$ as the time required for the SEF on the middle link to decrease to 30\% of its initial value. The choice of 30\% is made without loss of generality; a smaller fraction would require longer simulations but, due to the monotonic nature of thermalization, would not alter the results. This is not the thermalization time to reach the steady state but nevertheless facilitates the comparison between different system parameters that we vary, hence, it can be framed as a relative thermalization time. To exemplify this monotonicity of SEF from its initial value to zero, which allows for this definition of $\mathcal{T}$, we plot in figure~\ref{fig:monotonicity_of_sef} the SEF as a function of time for a set of parameters while emphasizing that all parameters follow the same pattern.

Regarding the parameters for this section we set $N = 12$ and $T = 10$. Further, for smaller $D$, we have less kinetic dissipation which allows the particles to reach the system boundaries within the thermalization time-frame~\cite{open_schwinger}. To minimize boundary effects, we constrain $D$ to the range $2 \leq D \leq 5$. Similarly, the applied background electric field is limited to $0 \leq l_0 \leq 0.5$ to avoid boundary effects and Bragg reflections~\cite{Su:2024uuc}. The condition $T \gg H_S$, required for the quantum Brownian motion regime, is verified numerically using ITensors' DMRG to compute the ground and first excited states of $H_S$~\cite{ITensors}. Specifically, the typical energy gap values for $N$ between 12 and 100, for masses between 0.1 and 1.0, and for $l_0$ between 0 and 0.5 are of the order of 1, which is an order of magnitude smaller than our chosen $T = 10$.

\begin{figure}
    \centering
    \includegraphics[width=\linewidth]{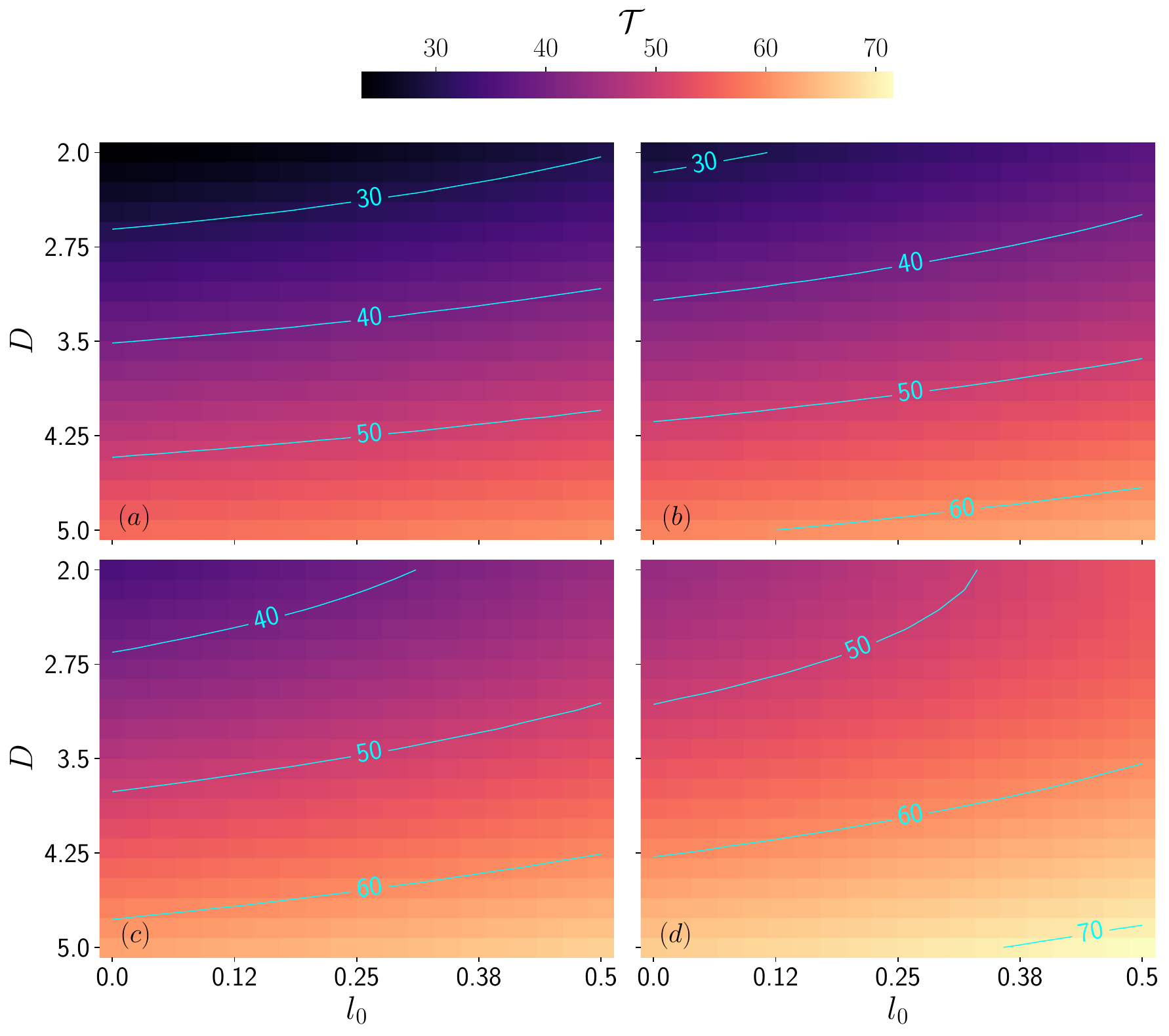}
    \caption{Thermalization time $\mathcal{T}$ as a function of the dissipator strength $D$ and the background electric field $l_0$. The parameters used are $N = 12$, $x = 1$, $T = 10$, $(a)$: $m = 0.1$, $(b)$: $m = 0.5$, $(c)$: $m = 0.75$, $(d)$: $m =1$, and each axis has 20 equidistant points with $D \in [2, 5]$, $l_0 \in [0.0, 0.5]$. All other parameters are set as mentioned in section~\ref{sec:results_discussion}. The cyan contour lines are levels for the thermalization time values $\mathcal{T}$. The thermalization time is defined as the time for the subtracted electric field on the middle link $\Delta F(n = 5)$ to reach 0.3 of its original value. Increasing $D$, $l_0$, $m$ gives a higher $\mathcal{T}$.}
    \label{fig:thermalization_times}
\end{figure}

\begin{figure}
    \centering
    \includegraphics[width=\linewidth]{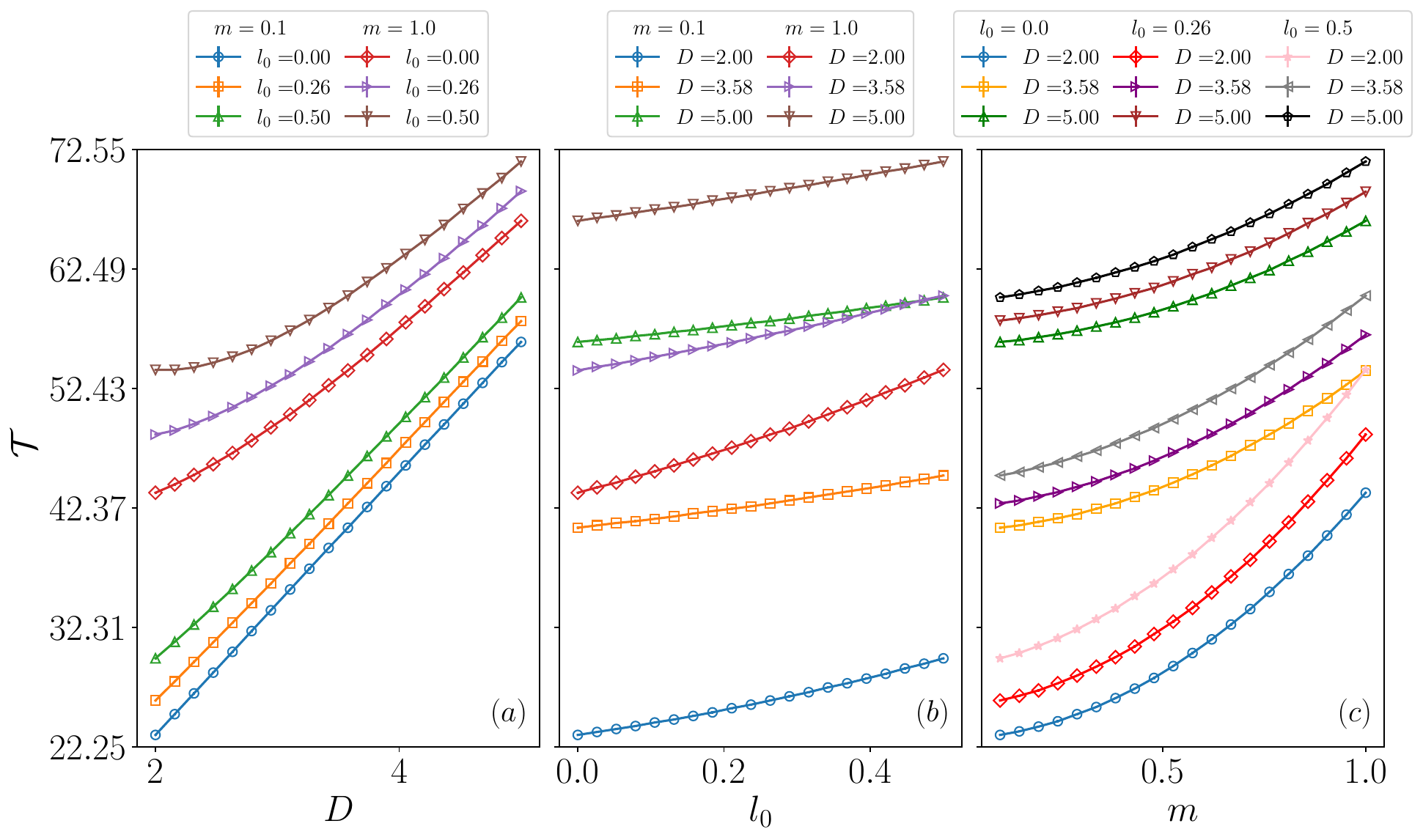}
    \caption{$(a)$: Thermalization time $\mathcal{T}$ vs dissipator strength $D$, $(b)$: applied background electric field $l_0$ and $(c)$: mass $m$. The error bars which are not visible due to the bigger y-scale of the plot are estimated to be $0.1$ as inferred from the data in table~\ref{tab:thermalization_times} and discussed in section~\ref{sec:larger_systems}. The lines are drawn solely for guidance.}
    \label{fig:thermalization_vs_individual}
\end{figure}

The results of the thermalization times $\mathcal{T}$ as a function of $D \in [2, 5]$, $l_0 \in [0, 0.5]$ are shown in figure~\ref{fig:thermalization_times} for masses $m = 0.1$, $0.5$, $0.75$, $1$. Additionally, in figure~\ref{fig:thermalization_vs_individual} we show the explicit dependence of $\mathcal{T}$ on each parameter separately.

\begin{figure}
    \centering
    \includegraphics[width=\linewidth]{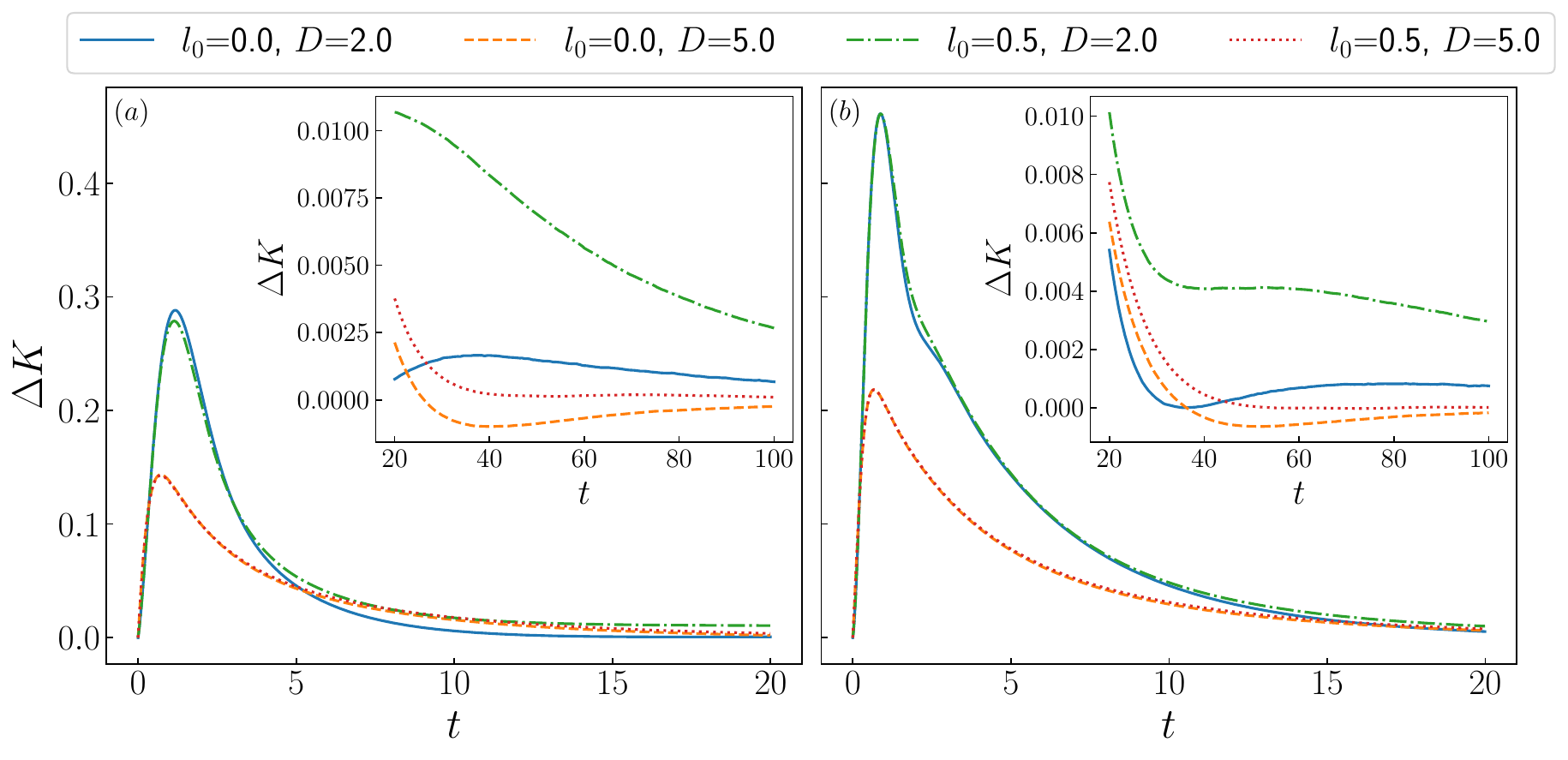}
    \caption{$(a)$: Subtracted kinetic energy $\Delta K$ vs time for $m = 0.1$ and $(b)$: $m = 1.0$ at $N = 12$. The resolution in $t$ is fine enough that the data is quasi-continuous and connected with lines. A higher dissipator strength $D$ leads to a smaller SKE maximum. The insets show a later stage of the time evolution towards thermalization into the steady state in greater detail.}
    \label{fig:KE_mass_0.1_1.0}
\end{figure}

The first deduction we can make from figure~\ref{fig:thermalization_times} is that the thermalization time increases with dissipator strength $D$. This persists for all masses and all values of $l_0$. In figure~\ref{fig:thermalization_vs_individual}$(a)$ the dependence of $\mathcal{T}$ on $D$ is shown to be linear for small $m$. At larger $m$, it is still linear for higher $D$, but at smaller $D$, a non-trivial small deviation can be observed which makes $\mathcal{T}$ less susceptible to changes in $D$ and which becomes more pronounced with increasing $l_0$. As aforementioned, a higher dissipator strength $D$ introduces more kinetic dissipation into the system~\cite{open_schwinger}, acting as a drag force, which causes the charges forming the string to move at lower speed and will thus slow down any inward contraction or outward expansion of the string. Numerical evidence for this is shown in figure~\ref{fig:KE_mass_0.1_1.0}, where we plot the subtracted kinetic energy (SKE) $\Delta K$ as a function of time for $m = 0.1, 1$ at $l_0 = 0, 0.5$ and $D = 2, 5$. Here we take the kinetic energy to correspond to the first line of Eq.~\eqref{eq:H_S}. In this plot for both $l_0 = 0, 0.5$ by increasing $D$ from 2 to 5 the SKE maximum for $m = 0.1$ goes from 0.28 to 0.14 and for $m = 1$ from 0.46 to 0.22. In other words, the height of the peak SKE is decreased for both masses as the dissipator is increased. From the insets we can also observe that a higher dissipator strength takes SKE faster to zero.

Another indication from figure~\ref{fig:thermalization_times} is that the thermalization time $\mathcal{T}$ increases with increasing applied background electric field $l_0$. This dependence is shown explicitly in figure~\ref{fig:thermalization_vs_individual}$(b)$ to have a behaviour which is slightly faster than linear. The background field is oriented in the direction of the initial string where the positive charge is to the left of the negative charge. Hence, according to the electrostatic force, the two original charges are squeezed together which slows down any outward moving expansion of the string.

Finally, as can be observed from figure~\ref{fig:thermalization_times} and figure~\ref{fig:thermalization_vs_individual}$(c)$, the thermalization time increases with increasing mass $m$. In figure~\ref{fig:thermalization_vs_individual}$(c)$ this increase of $\mathcal{T}$ with $m$ is shown to be faster than linear for the parameter regime we study. Moreover, at lower fixed $D$, the thermalization time $\mathcal{T}$ is more sensitive to changes in $m$. An increasing mass has the effect of slowing down the charges forming the string, as heavier charges move slower, which in turn slows down the string's contraction or expansion.

The common feature in increasing $D$, $l_0$, or $m$, is that they keep the charges closer together for a longer time period as the expanding component of the string is suppressed, thereby limiting the spatial extent of the wavefunction. This has a direct analogy to quarkonia moving and thermalizing in QGP. In this context, a spatially extended wavefunction decoheres faster~\cite{PhysRevD.101.034011, AKAMATSU2022103932}. In fact, decoherence is needed for dissociation of quarkonia in the process of thermalization~\cite{AKAMATSU2022103932}, hence, a delayed decoherence and dissociation leads to a delayed thermalization. Specifically for $D$, the classical picture is that this drag force will prevent the charge pair from dissociating~\cite{PhysRevD.101.034011}. For the case of increasing $m$, the bottomonium thermalizes slower than the charmonium as the bottomonium is heavier. In~\cite{PhysRevD.101.034011}, where this observation was made numerically using the quantum state diffusion method, it was noted that since the bottomonium, which is heavier than the charmonium, is more spatially localized, it decoheres slower prolonging the relaxation time. 

\subsection{Correlations between thermalization time and mutual information}
\label{sec:mutual_info}

Given the above discussion on dissociation, we continue for the same fixed parameters $N = 12, T = 10$ of the string case and investigate the effect of $D$, $l_0$, $m$ on the quantum mutual information which is defined by 
\begin{align}
\label{eq:mutual_info}
    &S(\rho) = -\mathrm{Tr}(\rho \ln \rho), \\
    &I(A, B) = S(\rho_{A}) + S(\rho_{B}) - S(\rho_{AB})
\end{align}
The first line defines the von Neumann entropy with $A$, $B$ being two separate contiguous subregions of the whole space, and the subscript on the density matrix implies that the rest of the region besides the subscript has been traced out. We track the subtracted mutual information (SMI) $\Delta I$ between the regions $A, B$ at lattice sites 4, 5 and 6, 7 respectively. As explained in section~\ref{sec:results_discussion}, the subtraction defining $\Delta I$ is between the observable value of the case when the initial state is the Dirac vacuum modified to include the pair of charges and the Dirac vacuum state itself. Using these two regions $A$ and $B$, we thus probe how one half of the string which includes the positive charge is correlated to the other half of the string which includes the negative charge. 

\begin{figure}
    \centering
    \includegraphics[width=\linewidth]{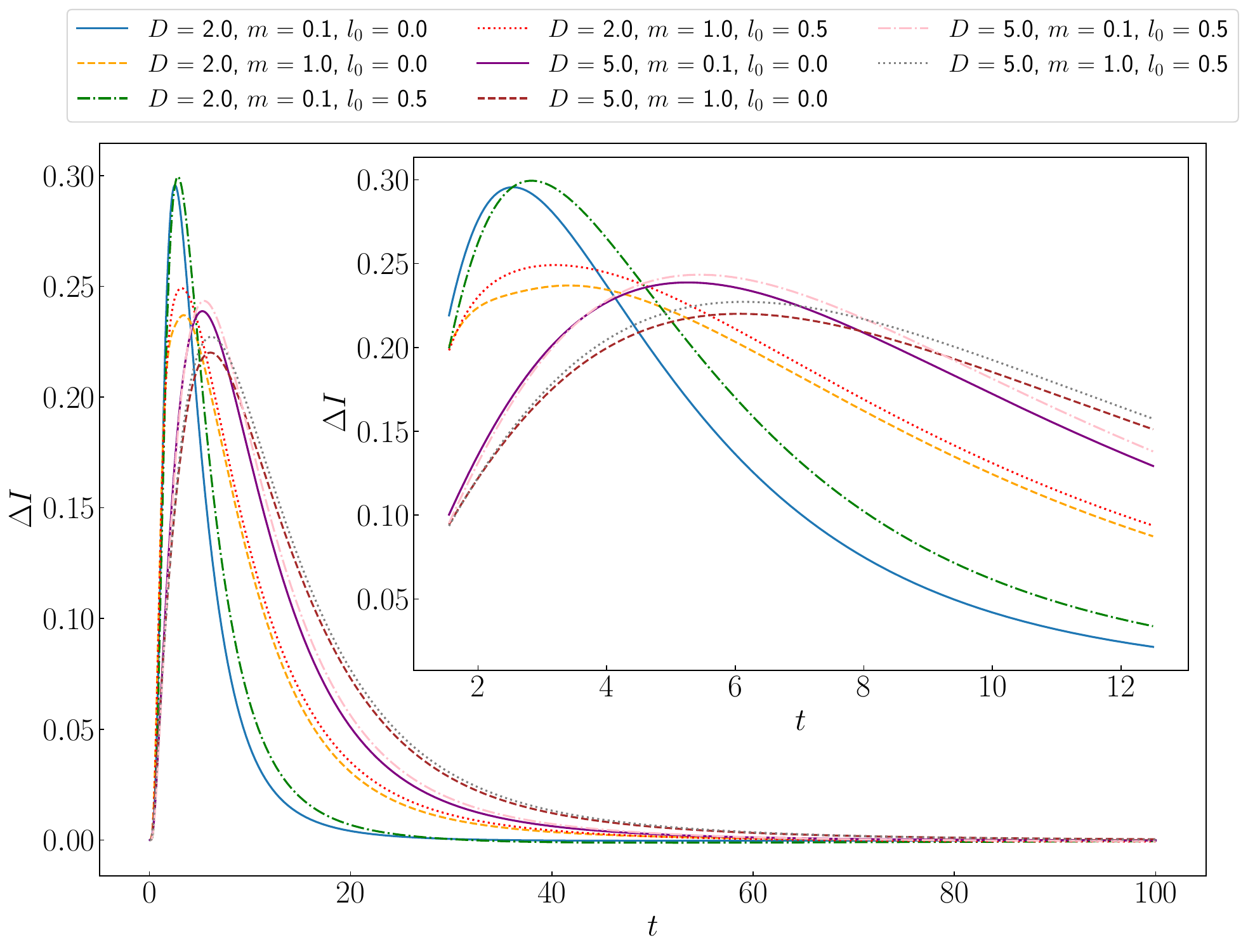}
    \caption{Subtracted mutual information (SMI) $\Delta I$ as a function of time $t$ for the parameter sets in the four corners of figure~\ref{fig:thermalization_times}(a, d) at $N = 12$. The mutual information is measured between sites 4,5 and 6,7. The initial string extends between sites 4 to 7, hence the two regions 4,5 and 6,7 measure the mutual information between the two halves of the string. The inset focuses on the times between 10 and 20 to emphasize how fast the SMI of each parameter set decreases. The resolution in $t$ is fine enough that the data is quasi-continuous and connected with lines.}
    \label{fig:mutual_info}
\end{figure}

The results are shown in figure~\ref{fig:mutual_info}. Consistent with the line of argument above regarding higher $D$, $m$, $l_0$ values slowing down dissociation, it is evident that increasing $D$, $m$ or $l_0$ gives a slower decrease of SMI towards $\Delta I = 0$ which is emphasized by the inset of the plot. The peak of the SMI formed during transient dynamics is increased with increasing $l_0$ and decreased with increasing $D$, $m$. This is because increasing $l_0$ is squeezing together the charges forming the string which keeps the state more localized, while increasing $D$, $m$ simply decreases their kinetic energy. We start from product states which do not contain any mutual information between any regions; a changing parameter that allows the charges to come closer together will build up more SMI from $\Delta I = 0$ during the evolution before decreasing back to $\Delta I = 0$ which is the steady state value. Further, a changing parameter that keeps the state more localized will slow down the decrease of $\Delta I$ to zero.

\subsection{Temperature dependence of the string's thermalization}
\label{sec:temperature}

The temperature $T$ is another important parameter in heavy-ion collision experiments, hence, this section is dedicated to exploring the effect of varying the temperature on the thermalization time $\mathcal{T}$ of the string at fixed $N = 12$. The temperatures explored do not fall lower than 10 to comply with the requirement from QBM that $T \gg H_S$, as discussed in section~\ref{sec:thermalization_times_vs_various_parameters}.

\begin{figure}
    \centering
    \includegraphics[width=\linewidth]{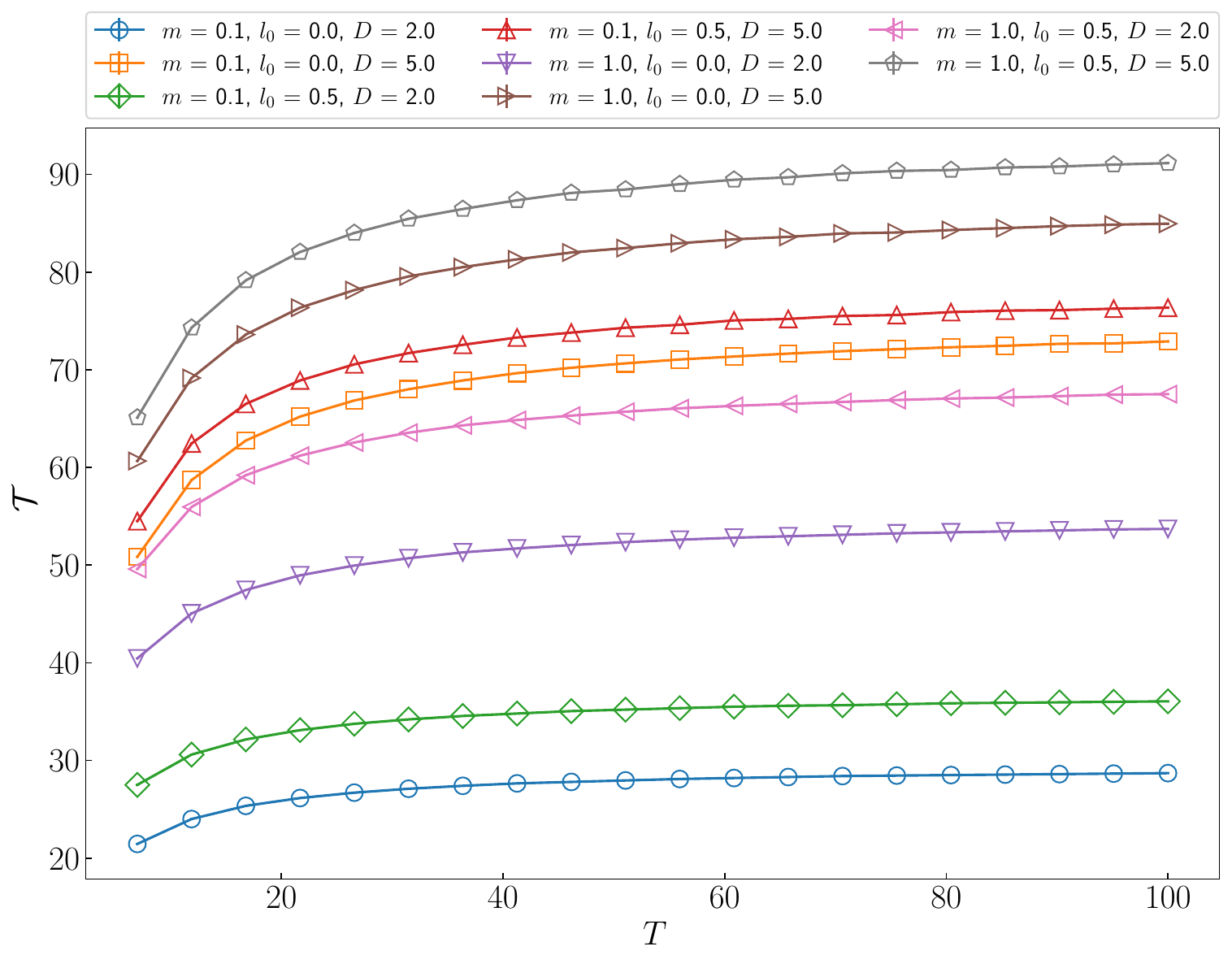}
    \caption{Thermalization time $\mathcal{T}$ as a function of the environment's temperature $T$ for $N = 12$, $D = 2, 5$, $l_0 = 0.0, 0.5$, $m = 0.1, 1.0$. The error bars, which are not visible due to the y-scale of the plot, are estimated to be $0.1$ as inferred from the data in table~\ref{tab:thermalization_times} and discussed in section~\ref{sec:larger_systems}. The lines represent the fit described by Eq.~\eqref{eq:fit_temperature}.}
    \label{fig:thermalization_time_vs_T}
\end{figure}

In figure~\ref{fig:thermalization_time_vs_T} we plot $\mathcal{T}$ as a function of $T$ for $D = 2, 5$, $l_0 = 0.0, 0.5$, and $m = 0.1, 1.0$. The conclusion from this figure is that the thermalization time increases with temperature. This is a feature of the Schwinger model that differs from the expectation of quarkonia in QGP~\cite{PhysRevD.110.074040}. For low temperatures we have a faster increase of $\mathcal{T}$ with increasing $T$. Above a certain temperature, we enter a linear regime. The figure also reassures the previous results in that at any fixed $T$ for a wide range of $T$, increasing $D, l_0$ or $m$ results in a larger $\mathcal{T}$.

In Eq.~(A3) of~\cite{open_schwinger}, the authors provide an approximation for the relaxation rate, which exhibits a quadratic dependence on the Lindblad jump operators in the numerator. Expanding this quadratic dependence results in a sum of terms proportional to \( 1/T^0 \), \( 1/T \), and \( 1/T^2 \). Consequently, the thermalization time is inversely proportional to this sum, directly explaining the observed increase in thermalization time with temperature in figure~\ref{fig:thermalization_time_vs_T}, as well as the initial nonlinear behavior and the transition to a linear regime. To quantitatively capture this dependence, we fit the data in figure~\ref{fig:thermalization_time_vs_T} using the function  
\begin{equation}
\label{eq:fit_temperature}
    f(T) = \frac{a}{(b+c/T)^2},
\end{equation}  
with fitting parameters \( a, b, c \). The fitted function accurately reproduces the observed functional behavior.  

In the case of QCD, the key difference arises from the temperature dependence of \( D(k) \). Specifically, \( D(k) \) scales as \( T^3 \), which leads to an increase in the relaxation rate with temperature. This behavior is due to the relationship between \( D(k) \) and the heavy quark diffusion coefficient \( \kappa \)~\cite{oqs_for_quarkonia, PhysRevD.97.074009}, which has been shown to exhibit a \( T^3 \) dependence~\cite{PhysRevLett.100.052301}.

\begin{figure}
    \centering
    \includegraphics[width=\linewidth]{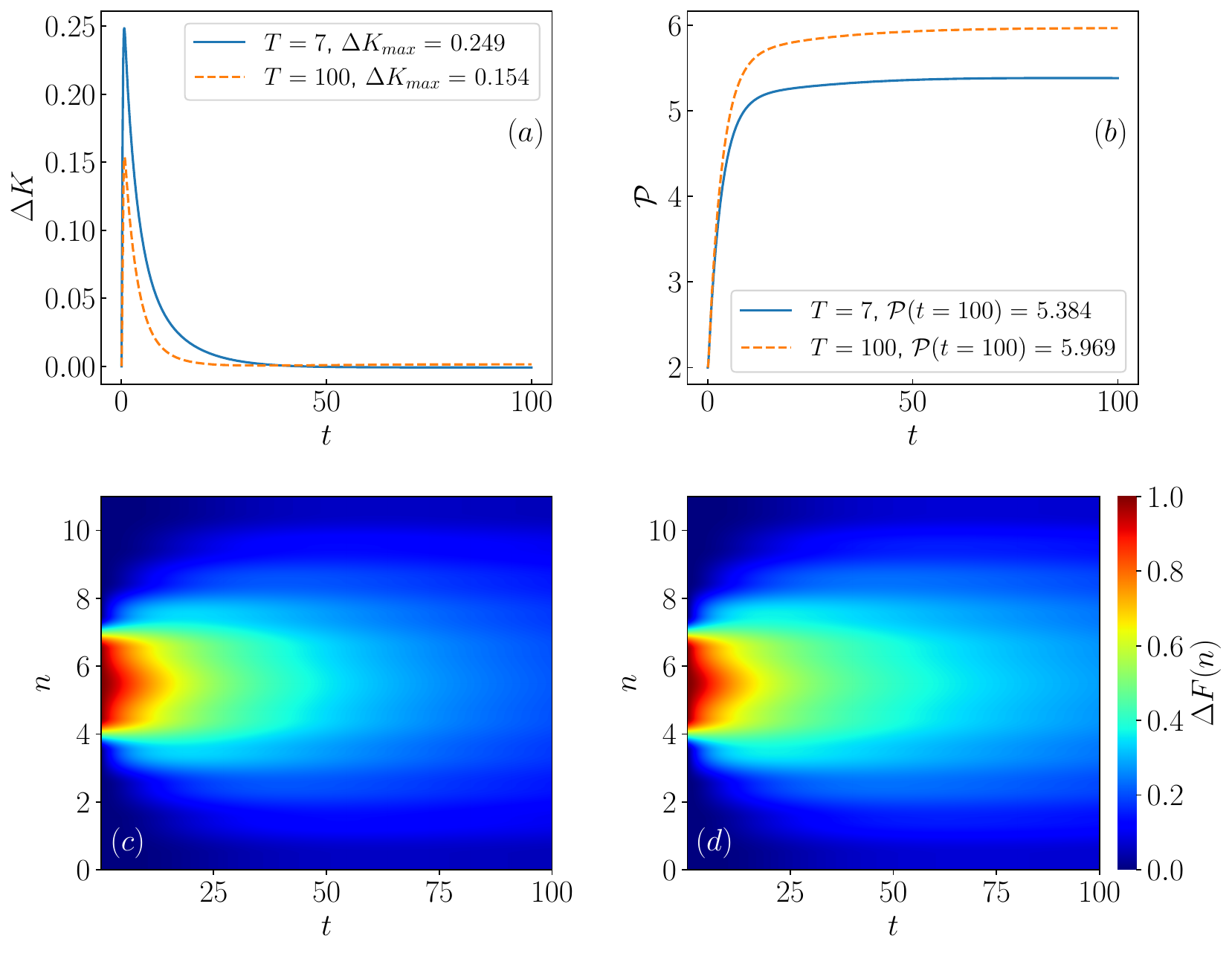}
    \caption{$(a)$: Subtracted kinetic energy $\Delta K$, $(b)$: particle number for the case where the initial state is the Dirac vacuum with the string present $\mathcal{P}$ $(b)$, both as a function of time $t$. In $(a)$ the legend gives the value of $\Delta K$ at the peak and in $(b)$ the legend gives the final value of $\mathcal{P}$ at time $t = 100$. The comparison is drawn between temperatures $T = 7$ and $T = 100$. The bottom two subplots $(c)$, $(d)$ show the subtracted electric field $\Delta F(n)$ as a function of the link number $n$ and time $t$ for $T = 7, 100$ respectively. All plots have fixed parameters $N = 12$, $D = 5$, $l_0 = 0.5$ and $m = 1.0$.}
    \label{fig:data_50}
\end{figure}

\begin{figure}
    \centering
    \includegraphics[width=\linewidth]{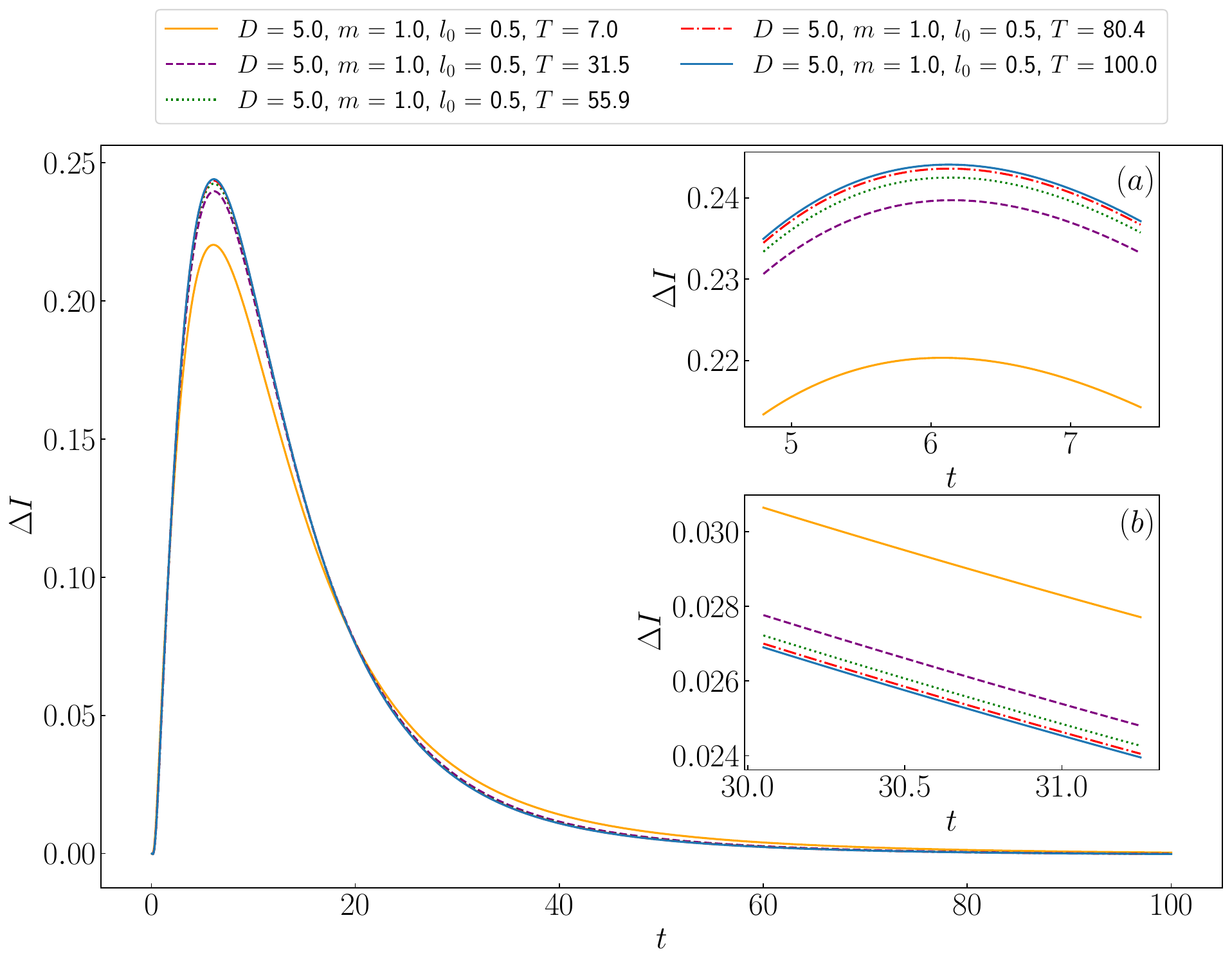}
    \caption{Subtracted mutual information $\Delta I$ as a function of time $t$ for $D = 5.0$, $m = 1.0$, $l_0 = 0.5$ and $T \in [7, 100]$. Inset $(a)$ focuses on the peak of $\Delta I$ and inset $(b)$ on the order at which different temperatures decrease to the steady state value $\Delta I = 0$ in late times.}
    \label{fig:mutual_info_with_T}
\end{figure}

In figure~\ref{fig:mutual_info_with_T} we present results for the subtracted mutual information $\Delta I$ as a function of time at $D = 5.0$, $m = 1.0$, $l_0 = 0.5$, for various temperature values over the range $T \in [7, 100]$. The figure shows an increasing peak value for $\Delta I$ with increasing $T$. Further, at late times, an increasing $T$ makes the $\Delta I$ tend to the steady state value faster. Both effects seem to be subtle but might be attributed to the following. Higher peaks at higher $T$ suggests that increasing $T$ allows the string's charges to build more mutual information in the transient time regime. The subtracted kinetic energy $\Delta K$ comparison between $T = 7$ and $T = 100$ shown in figure~\ref{fig:data_50}$(a)$ provides numerical evidence that higher $T$ gives a smaller $\Delta K$ peak, which in turn implies the charges might be spatially closer for a longer time as $T$ increases. Increasing $T$ also increases the particle number production that eventually can weaken the string tension through the creation of screening charges and destroy the mutual information faster. This would explain the late time pattern. Numerical evidence for the particle production levels are presented in figure~\ref{fig:data_50}$(b)$, where we plot the particle number for the case when the initial state is the Dirac vacuum with the string present to compare between $T = 7$ and $T = 100$. We call this observable $\mathcal{P}$ and define it mathematically as
\begin{equation}
\label{eq:particle_number_operator}
    \mathcal{P} = \frac{N}{2} + \frac{1}{2}\sum_{n=0}^{N-1}(-1)^nZ_n.
\end{equation}
The subfigures $(c)$, $(d)$ in figure~\ref{fig:data_50} show the subtracted electric field $\Delta F(n)$ as a function of the link number $n$ and time $t$ for $T = 7, 100$ respectively. From these we can qualitatively see that around $t = 50$ the links 3 and 8 for $T = 100$ have a larger $\Delta F(n)$ compared to $T = 7$ which implies the string expands more with larger $T$ at later times. This can be related to figure~\ref{fig:data_50}$(a)$ where the $T = 100$ case has a slower late time decrease to the steady state value of $\Delta K$ and this crossover occurs around $t = 50$, when $\Delta K$ for $T = 7$ goes below that of $T = 100$. Figure~\ref{fig:data_50}$(d)$ also shows the generally longer thermalization time that occurs with higher temperature as the middle links have a higher $\Delta F(n)$ for late times compared to figure~\ref{fig:data_50}$(c)$.

\subsection{Larger system sizes and symmetry preservation}
\label{sec:larger_systems}

To ensure our results are not affected by finite-size effects we present in table~\ref{tab:thermalization_times} quantitative results for various parameters, comparing the thermalization time at $N = 12$ and $N = 24$ at fixed $T = 10$. The table shows that the two agree to one decimal, from which we can thus also estimate an error for our thermalization time results to be of $\mathcal{O}\left(0.1\right)$.

\begin{figure}
    \centering
    \includegraphics[width=\linewidth]{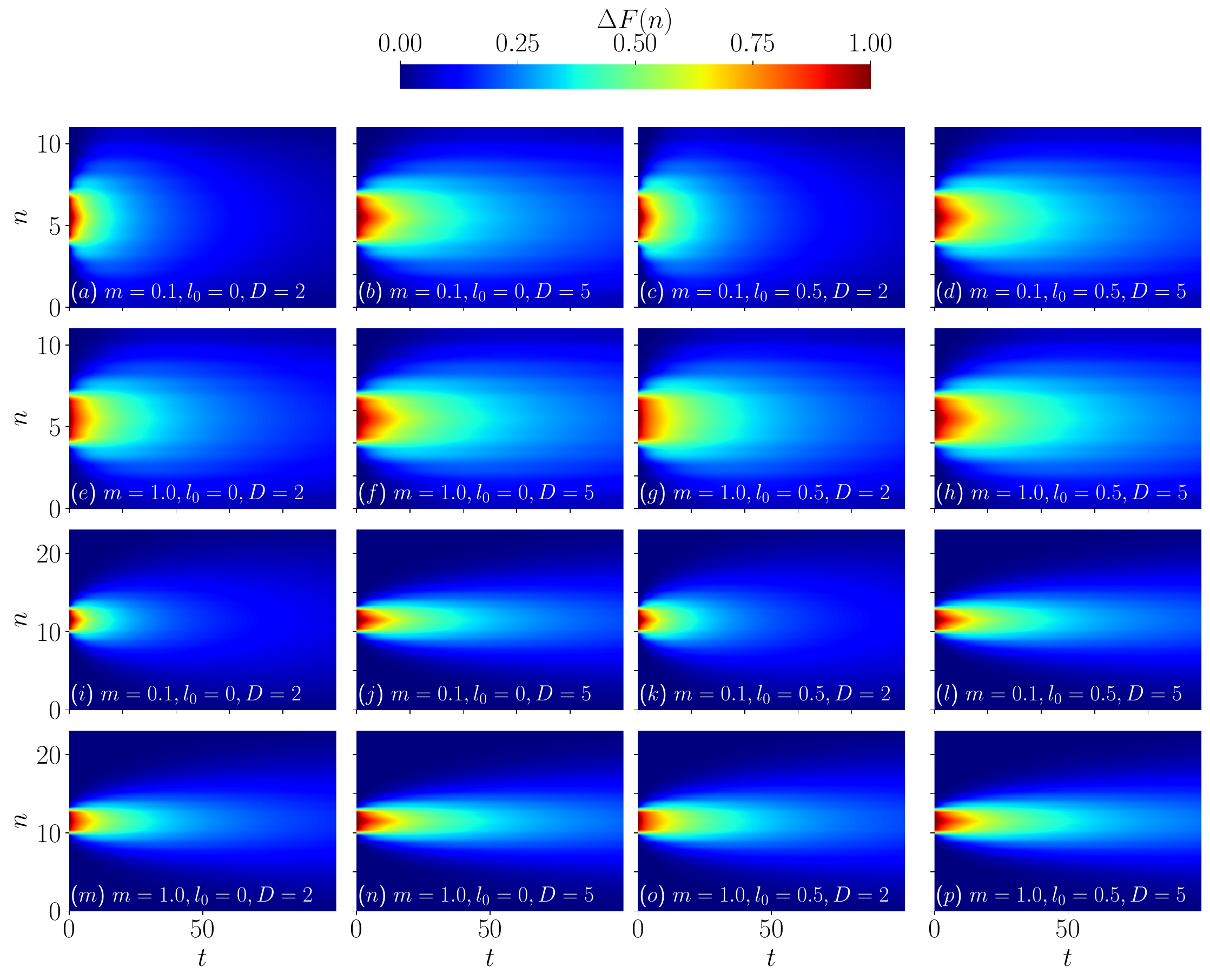}
    \caption{Subtracted electric field (SEF) $\Delta F(n)$ per link number $n$ as a function of time $t$ for $N = 12, 24$, $l_0 = 0.0, 0.5$, $D = 2.0, 5.0$, $m = 0.1, 1.0$. The rest of the parameters are as mentioned in section~\ref{sec:results_discussion}. The SEF should be zero when thermalized, hence, the plot indicates increasing thermalization time with increasing $D, l_0$ and $m$. By doubling the system size the evolution is kept further away from the boundaries.}
    \label{fig:ef_12_24}
\end{figure}

A further qualitative comparison between the two system sizes is given in figure~\ref{fig:ef_12_24}, where the SEF is plotted per link number for each time step. This figure exemplifies without loss of generality for the chosen parameters, how at $N = 24$, the dynamics behave similarly to $N = 12$, albeit with the $N = 24$ case better avoiding boundaries while maintaining good agreement with $N = 12$ in Table~\ref{tab:thermalization_times}. One can also observe directly from this figure how the SEF evolves and thermalizes to zero. For example, going from figure~\ref{fig:ef_12_24}$(a)$ to $(b)$ which takes $D$ from 2 to 5 at fixed $l_0 = 0$, $m = 0.1$, gives a more concentrated red region towards the middle link most likely because the outward speed of the charges is reduced from dissipation and the electrostatic attraction that is drawing the charges closer together. The effects of changing $l_0$ and $m$, besides the lengthening of the nonzero SEF region, are slightly more subtle to be observed distinctly in this qualitative figure.
\begin{table}[htbp]
    \centering
    \begin{tabular}{|c|c|c|c|c|c|}
        \hline
        \multirow{2}{*}{\centering $D$} & \multirow{2}{*}{\centering $l_0$} & \multicolumn{2}{c|}{$m$ = 0.1} & \multicolumn{2}{c|}{$m$ = 1.0} \\ 
        \cline{3-6}
        & & $N = 12$ & $N = 24$ & $N = 12$ & $N = 24$ \\ 
        \hline
        
        2.0 & 0.0 & 23.32 & 23.39 & 43.66 & 43.92 \\
        \hline
        3.58 & 0.0 & 40.89 & 41.14 & 54.14 & 53.93 \\
        \hline
        5.0 & 0.0 & 56.56 & 56.64 & 66.98 & 66.96 \\ 
        \hline

        2.0 & 0.26 & 26.15 & 26.63 & 48.68 & 49.26 \\
        \hline
        3.58 & 0.26 & 42.90 & 43.35 & 56.75 & 57.23 \\
        \hline
        5.0 & 0.26 & 58.20 & 58.78 & 68.86 & 69.25 \\ 
        \hline

        2.0 & 0.5 & 29.72 & 30.70 & 53.98 & 55.45 \\
        \hline
        3.58 & 0.5 & 45.10 & 45.77 & 59.99 & 60.90 \\
        \hline
        5.0 & 0.5 & 59.72 & 60.36 & 71.12 & 71.87 \\ 
        \hline
    \end{tabular}
    \caption{Thermalization times for different parameters at $N = 12$ and $N = 24$ with the rest of the parameter values given in section~\ref{sec:results_discussion}.}
    \label{tab:thermalization_times}
\end{table}

\begin{figure}
    \centering
    \includegraphics[width=\linewidth]{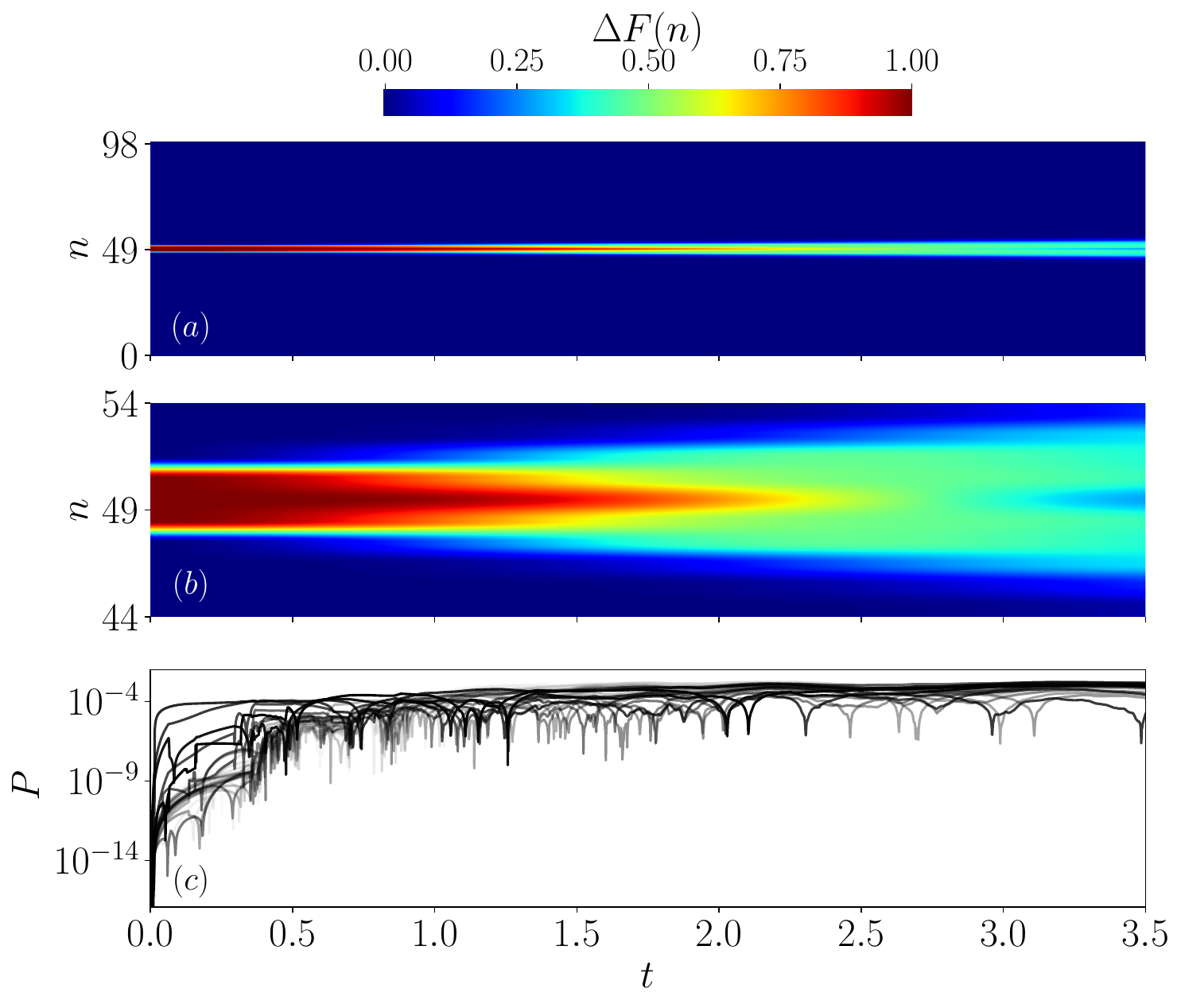}
    \caption{($a$): Subtracted electric field $\Delta F(n)$ per link number $n$ as a function of time $t$ for $N = 100$, $D = 0.15$, $x = 4$, $m = 0$, $\tau = 0.001$. ($b$) Focusing on the region around the middle links of subplot ($a$). ($c$): Absolute SEF difference between link pairs $P = \left|\Delta F(n = i) - \Delta F(n = N-i-2)\right|$, with $i \in [0, 48]$, paired by reflection across the middle link for the data in $(a)$. The evolution algorithm preserves the symmetry about the middle link to an average accuracy of $\mathcal{O}\left(10^{-4}\right)$. The darkest line is the pair of links next to the middle link and the brightest line is the pair of the first and last link.}
    \label{fig:ef_symmetry}
\end{figure}

The choice of dissipator $D(n-k) = D\delta_{n,k}$ is weakly $CP$-conserving. If we let $CP$ be the charge conjugation-parity operator, then this implies $[CP \otimes CP, \mathcal{L}] = 0$~\cite{neural_nets_oqs_schwinger}. With this choice of $D(n-k)$, the electric field is symmetric under reflection around the middle link~\cite{neural_nets_oqs_schwinger}. We present in figure~\ref{fig:ef_symmetry} how well our evolution algorithm preserves this symmetry. The parameters used are $N = 100$, $D = 0.15$, $x = 4$, $m = 0$, $\tau = 0.001$, with all the other parameters set as discussed in section~\ref{sec:methods}. These parameters were specifically chosen to match the ones used for figure 8$(a)$ in~\cite{neural_nets_oqs_schwinger} where a method of neural networks was explored on a smaller system size of $N = 20$. We plot in figure~\ref{fig:ef_symmetry}($c$) the absolute difference between the SEF at the first and last link, the second and penultimate link and so on. We thus define this observable as $P = \left|\Delta F(n = i) - \Delta F(n = N-i-2)\right|$, with $i \in [0, 48]$. From the figure we can see that our algorithm has preserved this symmetry to an average accuracy of $\mathcal{O}\left(10^{-4}\right)$, with the numerical average over all links being $P_\text{avg} = 0.0006$. The corresponding SEF per link as a function of time is given in figure~\ref{fig:ef_symmetry}($a$, $b$). It agrees with the qualitative behaviour seen in~\cite{neural_nets_oqs_schwinger}, albeit with qualitatively better performance in maintaining the reflection symmetry around the middle link, overall accuracy and stability of the evolution even for $N = 100$. The figure also forms an example of string breaking before thermalization.

\subsection{Schwinger boson thermalization}
\label{sec:schwinger_boson_results}

\begin{figure}
    \centering
    \includegraphics[width=\linewidth]{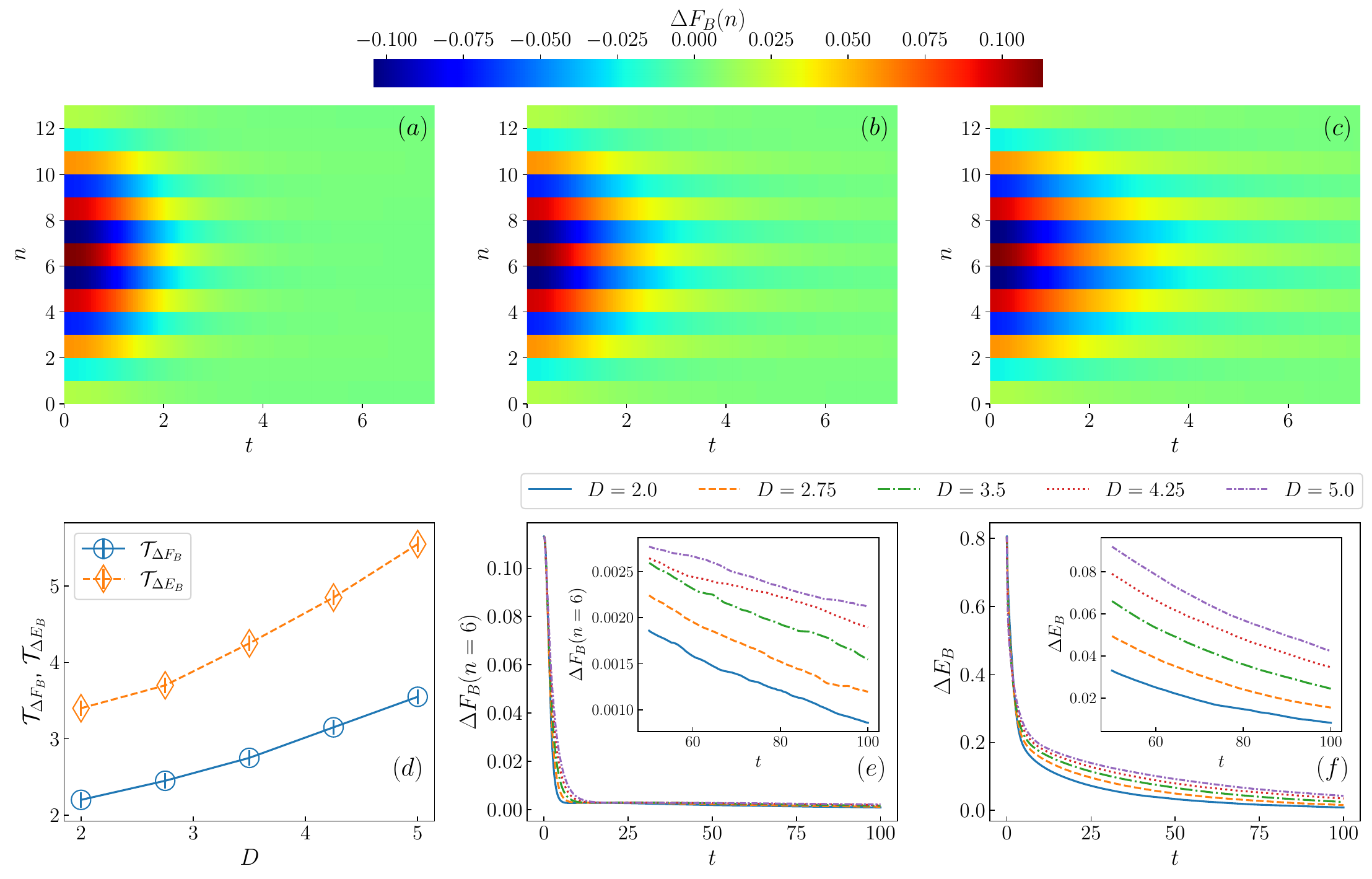}
    \caption{$(a)-(c)$: Subtracted electric field (SEF) of the Schwinger boson $\Delta F_B(n)$ per link number $n$ as a function of time $t$ for $D = 2, 3.5, 5$ respectively. $(d)$: Thermalization time as measured using the (SEF) $\mathcal{T}_{\Delta F_B}$ and subtracted energy (SE) of the Schwinger boson $\mathcal{T}_{\Delta E_B}$ vs the dissipator strength $D$. $(e)$, $(f)$: SEF, SE against time respectively for various values of $D$. The insets focus on the late time behaviour. In $(e)$, $(f)$, we omit plotting the individual points of the lines as they are spaced finely in time.}
    \label{fig:plot_57}
\end{figure}

This final section of the results discusses the Schwinger boson case as introduced in section~\ref{sec:results_discussion}. This section is important as it forms an independent test on the pattern of results observed for the string case, specifically for the dependence of the thermalization time as a function of the dissipator strength. The states involved in this section are entangled from the beginning, and thus show more complex dynamics. The first set of results are presented in figure~\ref{fig:plot_57}$(a-f)$ fixing $N = 14$, $T = 10$, $l_0 = 0$ and $m = 0$. In $(a-c)$ the subplots show the subtracted electric field of the Schwinger boson case, labeled $\Delta F_B(n)$, per link number $n$ and as a function of time $t$ for $D = 2, 3.5, 5$ respectively. It can be qualitatively observed that as $D$ is increased, the thermalization time increases, as calculated from the electric field of the middle link at $n = 6$. We thus confirm numerically that the pattern of results observed for the string case persists also for the case of the Schwinger boson. In subplot $(d)$ we plot the thermalization time as measured from $\Delta F_B(n = 6)$ and from the subtracted total energy $\Delta E_B$ of $H_S$. The subtracted total energy is used here as a crosscheck to the behaviour of $\mathcal{T}$ with $D$ when measured from the SEF. Thus $\mathcal{T}_{\Delta F_B}$ and $\mathcal{T}_{\Delta E_B}$ are plotted as a function of the dissipator strength $D$. The subplot shows that both observables give the same pattern of increasing thermalization time with increasing $D$, however, $\mathcal{T}_{\Delta E_B}$ gives consistently higher values. In subplots $(e)$, $(f)$ we plot the corresponding $\Delta F_B(n = 6)$ and $\Delta E_B$ as a function of time. The insets focus on the long-time behaviour, where the Trotterization error is larger, thus explaining the lack of smoothness in the lines. Nevertheless, the insets emphasize once again that a higher $D$ gives a slower observable thermalization.

\begin{figure}
    \centering
    \includegraphics[width=\linewidth]{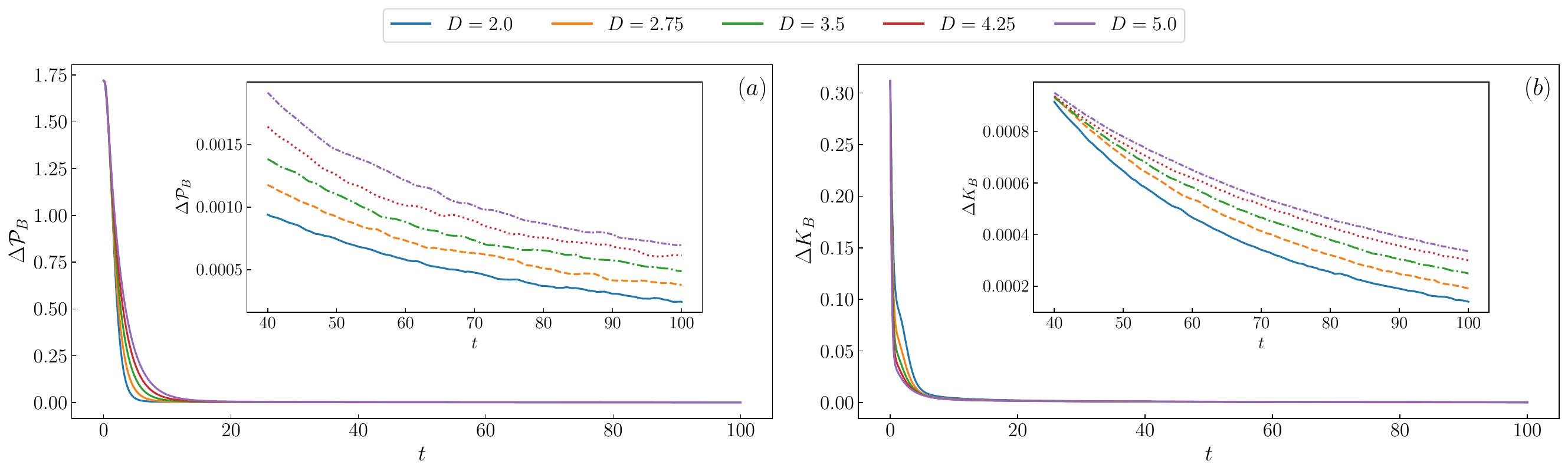}
    \caption{$(a)$: Subtracted particle number of the Schwinger boson $\Delta \mathcal{P}_B$ as a function of time $t$. $(b)$: Subtracted kinetic energy of the Schwinger boson $\Delta K_B$ vs time. The plots show various values of the dissipator strength $D$ and the insets focus on the late time behaviour. We omit plotting the individual points of the lines as they are spaced finely in time.}
    \label{fig:plot_57_pn_ke}
\end{figure}

Another set of important observables for the Schwinger boson case are presented in figure~\ref{fig:plot_57_pn_ke}. In subplot $(a)$, we plot the subtracted particle number $\Delta \mathcal{P}_B$ against time $t$, where the particle number $\mathcal{P}_B$ is defined as the expectation value of the operator in Eq.~\eqref{eq:particle_number_operator}. As expected, the values in the subplot start close to 2, since the Schwinger boson is a meson consisting of two particles bound together. In subplot $(b)$ the subtracted kinetic energy $\Delta K_B$ is plotted as a function of time. The important observation from both of these subplots is that a higher $D$ results in a slower thermalization rate for the observables.

\section{Conclusion and outlook}
\label{sec:conclusion}

In this work, we have explored the thermalization of a string formed by two opposite charges within the system represented by the lattice Schwinger model as well as of the Schwinger boson. These states evolve under the influence of a hot environment in thermal equilibrium represented by a $\phi^4$-theory. To derive the Lindblad master equation determining the evolution, we have made use of the Markovian and quantum Brownian motion limits. Using a matrix product state ansatz to represent the reduced density matrix of the open lattice Schwinger model and the adaptive time-dependent DMRG, we were able to perform time simulations and analyze the dependence of thermalization on various relevant parameters.

The electric field observable on the middle lattice link was used to define a thermalization time $\mathcal{T}$. A similar definition using the total energy of the system in the case of the Schwinger boson has shown a consistent pattern of results for the dependence of the thermalization time on $D$. Using the electric field definition, we have observed for both the string and the Schwinger boson initial states that $\mathcal{T}$ increases with increasing dissipator strength $D$, which determines the interaction strength between the system and the environment. Further, for the case of the string, $\mathcal{T}$ has also been observed to increase with increasing applied background electric field $l_0$, fermion mass $m$, and environment temperature $T$.

An analysis of quantum mutual information between the two halves of the string has provided further insight into the thermalization process. It has been shown that a longer thermalization time is positively correlated with a slower decrease of QMI when the temperature is fixed. In this setting, it was further demonstrated that the peak QMI increases when the system parameters favour the contraction of the string. For the case of changing temperature at fixed $D, l_0$ and $m$, we have found that a smaller $T$, which leads to a faster thermalization time, results in a lower QMI peak and a slower rate of QMI decrease.

Finally, using the string case, we have performed larger system size simulations to probe finite-size effects from which we have confirmed minor discrepancies in the measurements of $\mathcal{T}$. This has allowed us to estimate an error on $\mathcal{T}$ to $\mathcal{O}\left(0.1\right)$, which does not alter our conclusions made on the pattern of $\mathcal{T}$ for the parameter regime we study. An order of magnitude larger system size of 100 sites has also been successfully simulated to exemplify the ability of our choice of algorithm to scale to larger system sizes, necessary in lattice high-energy physics to complement experimental results. This demonstration has also shown the ability of our algorithm to maintain symmetries of the system such as the parity symmetry of the electric field to a precision of $\mathcal{O}\left(10^{-4}\right)$.

A possible extension to our algorithm can be a modification that would guarantee positivity of the density matrix throughout the time evolution~\cite{PhysRevLett.116.237201}, as checking this property explicitly is exponentially difficult~\cite{PhysRevLett.113.160503}. However, several examples of our results such as figure~\ref{fig:ef_symmetry}, have shown good performance without this guarantee. Further, in~\cite{PhysRevLett.114.220601} they argue that since $\mathcal{L}$ is a completely positive map, the map should have a fixed point, such that, reducing systematic errors should systematically lead towards a positive tensor network ansatz representing the density matrix.

Another potential to build upon this work is to probe different initial states or even theories that take the system closer to the goal of representing lattice QCD. For example, the development of higher dimensional tensor networks can assist the effort of simulating 2+1 QED~\cite{PhysRevX.10.041040}. Different limits such as the quantum optical limit, which is also relevant to quarkonia in QGP~\cite{oqs_for_quarkonia}, can also be explored in future work. 

Finally, we believe our findings may be relevant to high-energy theory and experiments, including those at the Large Hadron Collider or the Relativistic Heavy Ion Collider, which could further explore the impact of an applied electric field on the thermalization time of quarkonia in QGP. 

\acknowledgments
We thank Xiaojun Yao for the helpful discussions and many useful comments. This work is partly funded by the European Union’s Horizon 2020 Research and Innovation Programme under the Marie Sklodowska-Curie COFUND scheme with grant agreement no.\ 101034267.
This work is funded by the European Union’s Horizon Europe Framework Programme (HORIZON) under the ERA Chair scheme with grant agreement no.\ 101087126.
This work is supported with funds from the Ministry of Science, Research and Culture of the State of Brandenburg within the Centre for Quantum Technologies and Applications (CQTA). G.M. acknowledges support from INFN
through the projects “QUANTUM” and “NPQCD", from the Italian funding within the “Budget MUR - Dipartimenti di Eccellenza 2023–2027” - Quantum Sensing and Modelling for One-
Health (QuaSiModO), from PNRR MUR project CN00000013-“Italian National Centre on HPC, Big Data and Quantum Computing”, from the University of Bari via the 2023-UNBACLE-0244025 grant.
\begin{center}
    \includegraphics[width = 0.08\textwidth]{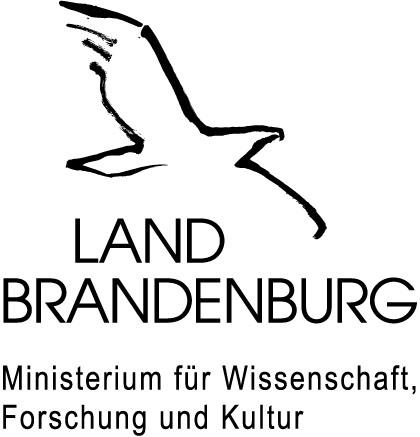}
\end{center}

\appendix

\section{Density Matrix MPS Representation}
\label{app:density_matrix_as_mps}

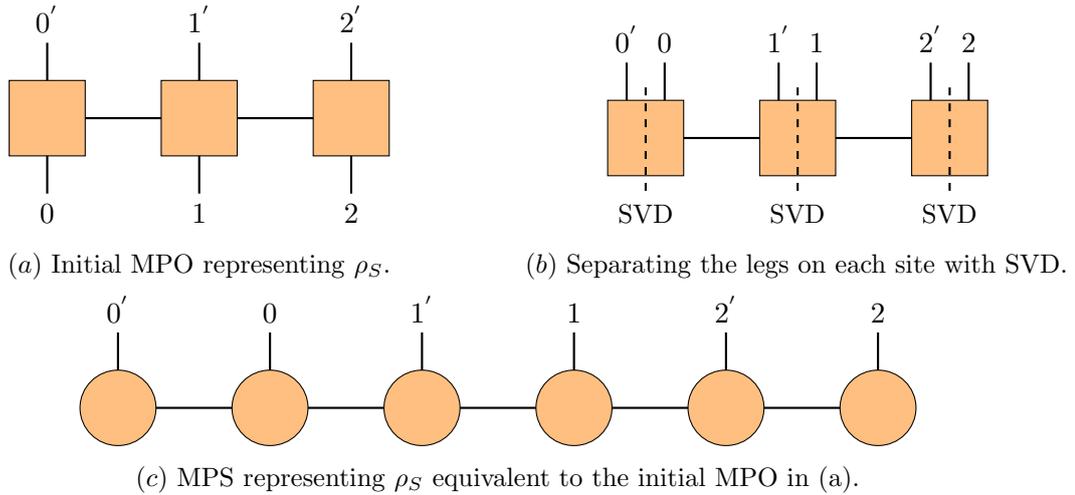
\begin{figure}[htbp]
\captionsetup[subfigure]{labelformat=empty}
    \centering
    \begin{subfigure}{0.48\textwidth}
        \centering
        \begin{tikzpicture}
            \foreach \x in {0, 2, 4} {
                \draw[fill=orange!50] (\x,0) rectangle ++(1,1);
                
                \draw[thick] (\x+0.5, 1) -- ++(0, 0.5);
                
                \draw[thick] (\x+0.5, 0) -- ++(0, -0.5);
            }
            
            Label indices for each site
            \foreach \x/\toplabel/\botlabel in {0/$0^{'}$/0, 2/$1^{'}$/1, 4/$2^{'}$/2} {
                \node[above] at (\x+0.5, 1.5) {\toplabel};
                \node[below] at (\x+0.5, -0.5) {\botlabel};
            }
            
            \foreach \x in {0, 2} {
                \draw[thick] (\x+1, 0.5) -- (\x+2, 0.5);
            }
        \end{tikzpicture}
        \caption{$(a)$ Initial MPO representing $\rho_S$.}
        \label{fig:mpo_diagram}
    \end{subfigure}
    \hfill
    \begin{subfigure}{0.48\textwidth}
        \centering
        \begin{tikzpicture}
            \foreach \x in {0, 2, 4} {
                \draw[fill=orange!50] (\x,0) rectangle ++(1,1);
                
                \draw[thick] (\x+0.75, 1) -- ++(0, 0.5);  
                \draw[thick] (\x+0.25, 1) -- ++(0, 0.5);  
                
                \draw[black, dashed, thick] (\x+0.5, -0.2) -- (\x+0.5, 1.2);
                \node[black, right] at (\x, -0.5) {\small SVD};
            }
            
            \foreach \x/\toplabel/\botlabel in {0/$0^{'}$/0, 2/$1^{'}$/1, 4/$2^{'}$/2} {
                \node[above] at (\x+0.25, 1.5) {\toplabel};  
                \node[above] at (\x+0.75, 1.5) {\botlabel};  
            }
            
            \foreach \x in {0, 2} {
                \draw[thick] (\x+1, 0.5) -- (\x+2, 0.5);
            }
        \end{tikzpicture}
        \caption{$(b)$ Separating the legs on each site with SVD.}
        \label{fig:mpo_with_svd}
    \end{subfigure}

    \begin{subfigure}{\textwidth}
        \centering
        \begin{tikzpicture}
            \foreach \x in {0, 2, 4, 6, 8, 10} {
                \draw[fill=orange!50] (\x,0) circle(0.5);
                
                \draw[thick] (\x, 0.5) -- ++(0, 0.5);  
            }
            
            \foreach \x/\toplabel in {0/$0^{'}$, 2/0, 4/$1^{'}$, 6/1, 8/$2^{'}$, 10/2} {
                \node[above] at (\x, 1.0) {\toplabel};  
            }
            
            \foreach \x in {0, 2, 4, 6, 8} {
                \draw[thick] (\x+0.5, 0) -- (\x+1.5, 0);
            }
        \end{tikzpicture}
        \caption{$(c)$ MPS representing $\rho_S$ equivalent to the initial MPO in (a).}
        \label{fig:mps_diagram}
    \end{subfigure}

    \caption{The transformation $(a) \to (c)$ from an MPO representing the system density matrix $\rho_S$ to an MPS through the application of SVD.}
    \label{fig:mpo_to_mps}
\end{figure}

The density matrix $\rho_S$ for which we simulate its time evolution to explore the dynamics generated by the Lindbladian equation of motion is an operator represented by a matrix product operator (MPO). The mathematical form for an MPO is given in Eq.~\eqref{eq:mpo_main_text}, which for the purpose of demonstration we present here again specifically for 3 sites
\begin{equation}
    \label{eq:mpo_equation_appendix}
    \rho_S = W_{\sigma_0\alpha_0}^{\sigma'_0}W_{\sigma_1\alpha_0\alpha_1}^{\sigma'_1}W_{\sigma_2\alpha_1}^{\sigma'_2}\ket{\sigma'_0\sigma'_1\sigma_2'}\bra{\sigma_0\sigma_1\sigma_2}.
\end{equation}
In figure~\ref{fig:mpo_to_mps} the procedure to transform this MPO to a matrix product state (MPS) is shown. Specifically, Eq.~\eqref{eq:mpo_equation_appendix} corresponds to figure~\ref{fig:mpo_to_mps}$(a)$ and the MPS equation for figure~\ref{fig:mpo_to_mps}$(c)$ is given by,
\begin{equation}
    \rho_S = A_{\sigma_0\alpha_0}A_{\sigma_1\alpha_0\alpha_1}A_{\sigma_2\alpha_1\alpha_2}A_{\sigma_3\alpha_2\alpha_3}A_{\sigma_4\alpha_3\alpha_4}A_{\sigma_5\alpha_4}\ket{\sigma'_0\sigma_0\sigma'_1\sigma_1\sigma_2'\sigma_2}.
\end{equation}
This has the advantage that the Taylor expansion in Eq.~\eqref{eq:Trotterization} can be applied with only one MPO to MPS contraction rather than applying an MPO to both sides of an MPO which would be the case if we kept the MPO representation of $\rho_S$. Through a simple procedure of singular value decompositions (SVD), the figure shows how $\rho_S$ is transformed into an MPS.

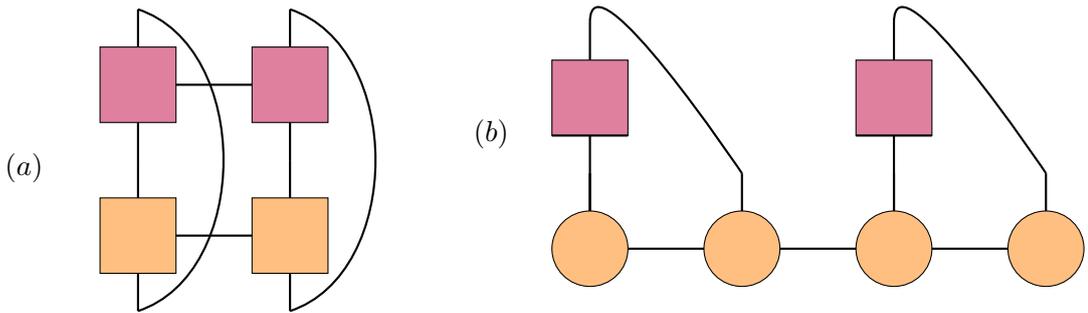
\begin{figure}[htbp]
    \centering
    \captionsetup[subfigure]{labelformat=empty} 

    \begin{subfigure}{0.49\textwidth}
        \begin{tikzpicture}
            \node at (-1, -0.6) {$(a)$};

            \foreach \x in {0, 2} {
                \draw[fill=purple!50] (\x, 0) rectangle ++(1, 1);
                \draw[thick] (\x+0.5, 1) -- ++(0, 0.5); 
                \draw[thick] (\x+0.5, 0) -- ++(0, -0.5); 
            }

            \foreach \x in {0, 2} {
                \draw[fill=orange!50] (\x, -2) rectangle ++(1, 1);
                \draw[thick] (\x+0.5, -2) -- ++(0, -0.5); 
                \draw[thick] (\x+0.5, -1) -- ++(0, 0.5); 
            }

            \foreach \x in {0} {
                \draw[thick] (\x+1, 0.5) -- (\x+2, 0.5); 
                \draw[thick] (\x+1, -1.5) -- (\x+2, -1.5); 
            }

            \draw[thick, black] (0.5, 1.5) .. controls (2, 1) and (2, -2) .. (0.5, -2.5);
            \draw[thick, black] (2.5, 1.5) .. controls (4, 1) and (4, -2) .. (2.5, -2.5);
        \end{tikzpicture}
    \end{subfigure}
    \hspace{-1.5cm} 
    \begin{subfigure}{0.49\textwidth}
        \begin{tikzpicture}
            \node at (-1.3, 1.52) {$(b)$};

            \foreach \x in {0, 2, 4, 6} {
                \draw[fill=orange!50] (\x, 0) circle(0.5);
                \draw[thick] (\x, 0.5) -- ++(0, 0.5); 
            }

            \foreach \x in {0, 2, 4} {
                \draw[thick] (\x+0.5, 0) -- (\x+1.5, 0);
            }
            \draw[thick] (0, 2) -- (4, 2);
            \foreach \x/\y in {0/0, 4/4} {
                \draw[fill=purple!50] (\x-0.5, 1.5) rectangle ++(1, 1);
                \draw[thick] (\x, 2.5) -- ++(0, 0.5);
                \draw[thick] (\x, 1.5) -- (\y, 0.5);
            }
        
            \draw[thick, black] (0, 3) .. controls (0, 4) and (2, 1) .. (2, 1);
            \draw[thick, black] (4, 3) .. controls (4, 4) and (6, 1) .. (6, 1);
            
            \foreach \x in {0, 4} {
                \draw[thick, black] (\x-0.5, 1.5) -- (\x+0.5, 1.5);
            }
        \end{tikzpicture}
        \vspace{-0.5em}
    \end{subfigure}

    \caption{This transformation from $(a)$ to $(b)$ shows how to measure observables in the MPS form of the density matrix. For a given observable operator $O$ represented by the purple MPO above the orange MPO, $(a)$ shows tr($O\rho_S$) and $(b)$ shows the same after $\rho_S$ has been converted to MPS.}
    \label{fig:measuring_observables_on_mps}
\end{figure}

We often need to measure the expectation value of an MPO operator $O$ with respect to the MPS representation of $\rho_S$. Figure~\ref{fig:measuring_observables_on_mps} shows graphically how this is done. In figure~\ref{fig:measuring_observables_on_mps}$(a)$ we have the case of an MPO to MPO contraction representing tr($O\rho_S$). The purple MPO on top is $O$ and the orange MPO on the bottom is $\rho_S$ in its original MPO form. In figure~\ref{fig:measuring_observables_on_mps}$(b)$ the figure shows how this is done with $\rho_S$ in its MPS form.

\section{Adaptive Time-Dependent DMRG}
\label{app:atddmrg}

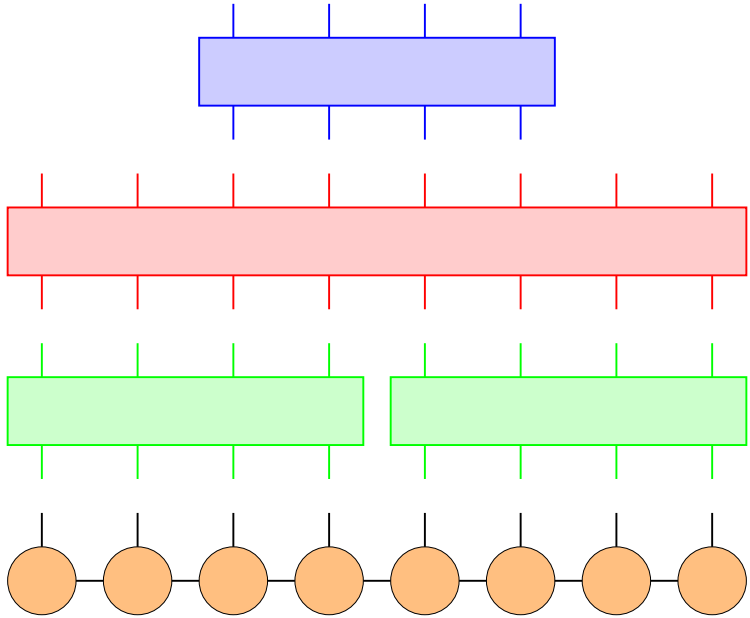
\begin{figure}[htbp]
\centering
\scalebox{0.9}{ 
\begin{tikzpicture}

    \def\spacing{1.4}  

    \foreach \x in {0, 1, 2, 3, 4, 5, 6, 7} {
        \draw[fill=orange!50] (\x * \spacing, -0.5) circle(0.5);

        \draw[thick] (\x * \spacing, 0) -- ++(0, 0.5);
    }

    \foreach \x in {0, 1, 2, 3, 4, 5, 6} {
        \draw[thick] (\x * \spacing + 0.5, -0.5) -- (\x * \spacing + \spacing - 0.5, -0.5);
    }

    \draw[thick, green, fill=green!20] (0 * \spacing - 0.5, 1.5) rectangle (3 * \spacing + 0.5, 2.5);
    \foreach \x in {0, 1, 2, 3} {
        \draw[thick, green] (\x * \spacing, 2.5) -- ++(0, 0.5); 
        \draw[thick, green] (\x * \spacing, 1.5) -- ++(0, -0.5); 
    }

    \draw[thick, green, fill=green!20] (4 * \spacing - 0.5, 1.5) rectangle (7 * \spacing + 0.5, 2.5);
    \foreach \x in {4, 5, 6, 7} {
        \draw[thick, green] (\x * \spacing, 2.5) -- ++(0, 0.5); 
        \draw[thick, green] (\x * \spacing, 1.5) -- ++(0, -0.5); 
    }

    % Creating individual red boxes and connecting them with horizontal lines
    \foreach \x in {0, 1, 2, 3, 4, 5, 6, 7} {
        \draw[thick, red, fill=red!20] (\x * \spacing - 0.5, 4) rectangle (\x * \spacing + 0.5, 5);
        \draw[thick, red] (\x * \spacing, 5) -- ++(0, 0.5); 
        \draw[thick, red] (\x * \spacing, 4) -- ++(0, -0.5); 
    }

    % Adding horizontal red connections between red boxes
    \foreach \x in {0, 1, 2, 3, 4, 5, 6} {
        \draw[thick, red] (\x * \spacing + 0.5, 4.5) -- (\x * \spacing + \spacing - 0.5, 4.5);
    }

    \draw[thick, blue, fill=blue!20] (2 * \spacing - 0.5, 6.5) rectangle (5 * \spacing + 0.5, 7.5);
    \foreach \x in {2, 3, 4, 5} {
        \draw[thick, blue] (\x * \spacing, 7.5) -- ++(0, 0.5); 
        \draw[thick, blue] (\x * \spacing, 6.5) -- ++(0, -0.5); 
    }

\end{tikzpicture}
} 
\caption{Schematic drawing example of the even group (bottom green boxes), Taylor group (middle red boxes) and odd group (top blue box) for $\mathcal{L}$ in Eq.~\eqref{eq:lindblad_operator_double_space_form} at $N = 4$. These groups are applied to the MPS at the very bottom representing $\rho_S$ as required by Eq.~\eqref{eq:Trotterization}, although here only $e^{\tau\mathcal{L}_O}e^{\frac{\tau}{2}\mathcal{L}_T}e^{\frac{\tau}{2}\mathcal{L}_E}$ is shown.}
\label{fig:odd_even_taylor_groups}
\end{figure}

The Lindblad operator $\mathcal{L}$ of Eq.~\eqref{eq:lindblad_operator_double_space_form} is partitioned in three groups which we call even, odd and Taylor as shown in figure~\ref{fig:odd_even_taylor_groups}. The even group (bottom green boxes) includes four-site operators that start at sites 0, 4 etc. The odd group (top blue box) includes four-site operators that start at sites 2, 6 etc. Any other term from $\mathcal{L}$ that does not fit into these two groups is put into the Taylor group. The exponentiation of this group as required in time evolution shown in Eq.~\eqref{eq:Trotterization} is done using the Taylor expansion approximation. We can then express the Taylor expansion operator as an MPO (middle red boxes).

The main idea of the adaptive time-dependent DMRG algorithm~\cite{Xiang_2023} is to maintain the mixed canonical form~\cite{SCHOLLWOCK201196} of the MPS after every multiplication onto the MPS such that the SVD truncation applied after every multiplication of a four-site tensor or the global Taylor MPO is optimal. Hence we start with the even group and the first multiplication is of the four-site tensor going from site 0 to site 3. This is done on an MPS in right canonical form~\cite{SCHOLLWOCK201196}. Next we have the multiplication of the second green four-site tensor going from site 4 to site 7. Before this, we put the MPS on sites 0 to 3 in left canonical form~\cite{SCHOLLWOCK201196} via QR decomposition with no truncation. Before the application of the global Taylor MPO there is no need to ensure any canonical form. After the global Taylor MPO comes the first four-site tensor of the odd group that starts from site 2 and ends at site 5. This means we set the MPS sites 0, 1 to left and 6, 7 to right canonical form before its application. We continue in the same pattern to finish with all four-site tensors in the odd group (when the system is larger than this example) and in general maintain the same procedure for the whole of Eq.~\eqref{eq:Trotterization}. This completes a single time step which we repeat to reach a desired total time $t$.

\bibliographystyle{JHEP}
\bibliography{biblio.bib}

\end{document}